\documentclass[lettersize,journal]{IEEEtran}
\usepackage{amsmath,amsfonts}
\usepackage{algorithmic}
\usepackage{algorithm}
\usepackage{array}
\usepackage[caption=false,font=normalsize,labelfont=sf,textfont=sf]{subfig}
\usepackage{textcomp}
\usepackage{stfloats}
\usepackage{url}
\usepackage{verbatim}
\usepackage{graphicx}
\usepackage{cite}
\usepackage{color}
\usepackage{amsthm}
\usepackage{booktabs}
\usepackage{multirow}
\usepackage{amssymb}
\usepackage{xspace}
\usepackage{enumitem}
\usepackage{xcolor}
\def\BibTeX{{\rm B\kern-.05em{\sc i\kern-.025em b}\kern-.08em
    T\kern-.1667em\lower.7ex\hbox{E}\kern-.125emX}}
\usepackage{balance}

\newcommand{\ignore}[1]{}

\newcommand{\ie}{\emph{i.e., }}

\newcommand{\eg}{\emph{e.g., }}

\newcommand{\citep}{\cite}
\usepackage{etoolbox}
\usepackage{flushend}

\begin{document}

\title{A Survey on Reinforcement Learning for Recommender Systems}


\author{
    Yuanguo Lin$^{1}$,
    Yong Liu$^{1}$,
    Fan Lin$^{*}$,
    Lixin Zou$^{*}$,
    Pengcheng Wu,
    Wenhua Zeng,
    Huanhuan Chen,
    Chunyan Miao

    \IEEEcompsocitemizethanks{
        \IEEEcompsocthanksitem Y. Lin is with the School of Computer Engineering, Jimei University; and the School of Informatics, Xiamen University, China. Email: xdlyg@stu.xmu.edu.cn.
        \IEEEcompsocthanksitem F. Lin and W. Zeng are with the School of Informatics, Xiamen University, China. Email: iamafan@xmu.edu.cn and whzeng@xmu.edu.cn.
        \IEEEcompsocthanksitem Y. Liu and P. Wu are with the Joint NTU-UBC Research Centre of Excellence in Active Living for the Elderly (LILY), Nanyang Technological University, Singapore. Email: stephenliu@ntu.edu.sg and pengchengwu@ntu.edu.sg.
        \IEEEcompsocthanksitem C. Miao is with the School of Computer Science and Engineering, Nanyang Technological University, Singapore. Email: ASCYMiao@ntu.edu.sg.
        \IEEEcompsocthanksitem L. Zou is with the School of Cyber Science and Engineering, Wuhan University, China. Email: zoulixin@whu.edu.cn.
        \IEEEcompsocthanksitem H. Chen is with the School of Computer Science and Technology, University of Science and Technology of China, China. Email: hchen@ustc.edu.cn.
        \IEEEcompsocthanksitem $^{*}$ Corresponding author
        \IEEEcompsocthanksitem $^{1}$ Co-first authors
    }


}
\markboth{IEEE TRANSACTIONS ON NEURAL NETWORKS AND LEARNING SYSTEMS, Accepted 17 May 2023.}%
{Shell \MakeLowercase{\textit{et al.}}: A Sample Article Using IEEEtran.cls for IEEE Journals}


\maketitle

\begin{abstract}
Recommender systems have been widely applied in different real-life scenarios to help us find useful information. In particular, Reinforcement Learning (RL) based recommender systems have become an emerging research topic in recent years, owing to the interactive nature and autonomous learning ability. Empirical results show that RL-based recommendation methods often surpass supervised learning methods. Nevertheless, there are various challenges of applying RL in recommender systems. To understand the challenges and relevant solutions, there should be a reference for researchers and practitioners working on RL-based recommender systems. To this end, we firstly provide a thorough overview, comparisons, and summarization of RL approaches applied in four typical recommendation scenarios, including interactive recommendation, conversational recommendation, sequential recommendation, and explainable recommendation. Furthermore, we systematically analyze the challenges and relevant solutions on the basis of existing literature. Finally, under discussion for open issues of RL and its limitations of recommender systems, we highlight some potential research directions in this field.
\end{abstract}


\section{Introduction}

\IEEEPARstart{P}{ersonalized} recommender systems~\cite{Deng198, Bobadilla115, Huang199} are competent to provide interesting information that matches users' preferences, and thereby alleviating the information overload problem. 
Recommendation technologies usually make use of various information to provide potential items for users. To this end, the early recommendation research primarily focuses on developing content-based and collaborative filtering-based methods~\cite{Adomavicius117,Shi118}. Recently, motivated by the quick developments of deep learning, various neural recommendation methods have been developed~\cite{Zhang119}. However, modeling the various information is not enough. In real-world scenarios, the recommender system suggests items according to the user-item interaction history, and then receives user feedback to make further recommendations~\cite{Zhao03, Pan34}. In other words, the recommender system aims to obtain users’ preferences from the interactions and recommend items that users may be interested in. Nevertheless, existing recommendation methods (\eg supervised learning) usually ignore the interactions between a user and the recommendation model. They do not effectively capture the user's timely feedback to update the recommendation model, thus leading to unsatisfactory recommendation results.

In general, the recommendation task could be modeled as an interactive process, \ie the user is recommended an item and then provides feedback (\eg skip, click or purchase) for the recommendation model. In the next interaction, the recommendation model learns the optimal policy from the user's explicit/implicit feedback and recommends a new item to the user. From the user's point of view, an efficient interaction means helping users find their favorite items as soon as possible.
The interactive recommendation approach has been applied in real-world recommendation tasks. However, it often suffers from some problems, \eg cold-start \cite{Huang58, Ji193}, data sparsity \cite{Zhao30}, interpretability \cite{Wang07} and safety \cite{Cao40}.

As a machine learning method that focuses on how an intelligent agent interacts with its environment, Reinforcement Learning (RL)~\cite{Neftci01, Li197} learns the policy by trial and error search, which is beneficial to sequential decision making. Hence, it can provide potential solutions to model the interactions between the user and agent. In particular, Deep Reinforcement Learning (DRL)~\cite{Kai113}, the combination of traditional RL with deep learning methods, is competent to learn from historical data with enormous state and action spaces to address large-scale problems. It has powerful representation learning and function approximation properties to be applied across various fields \cite{Zheng09, Zha203}, \eg games \cite{Max63} and robotics \cite{Kober106}. 
Recently, the application of RL to solve recommendation problems has become a new research trend in recommender systems~\cite{Zou36, Liu92, Ji01}. Specifically, RL enables the recommender agent to constantly recommend items to users for learning the optimal recommendation policies. Many experimental results have demonstrated that RL-based recommendation methods~\cite{Lee213, Wang223} evidently outperform supervised learning methods. For example, as shown in Table~\ref{EX}, RL-based recommendation methods (\ie PGPR~\cite{Xian06}, Actor-Critic~\cite{Lillicrap227}, and ADAC~\cite{Zhao57}) consistently perform better than the supervised learning-based recommendation methods (\ie DKN~\cite{Wang01}, BPR~\cite{Rendle01}, and RippleNet~\cite{Wang226}) on two Amazon datasets in terms of Hit Ratio (HR) and Normalized Discounted Cumulative Gain (NDCG), especially with significant margin (p$-$value $<$ 0.01) on the Clothing dataset.
In practice, RL-based recommender systems have been applied to many specific scenarios \cite{Zhang180, Zhou17, Fu183, Sun172, Wang221, Wei151}, such as e-commerce~\cite{He22, Ke132, Jangmin225}, e-learning~\cite{Zhang29, Lin205}, and health care~\cite{Wang32, Zheng181, Susan220}. 
\begin{table*}[t]
	\centering
	\caption{The recommendation performance of supervised learning methods (\ie DKN, BPR, RippleNet) and RL-based methods (\ie PGPR, Actor-Critic, ADAC) on two Amazon datasets in terms of HR and NDCG (\%). The p$-$value denotes the t-test used to test the performance differences between PGPR and other methods.}
	\label{EX}
\begin{tabular}{lllllll}
\hline
Dataset & \multicolumn{3}{c}{Beauty} & \multicolumn{3}{c}{Clothing} \\ \hline
Metrics & HR      & NDCG    & p$-$value & HR      & NDCG   & p$-$value  \\ \hline
DKN      & 8.673$\pm$0.058   & 1.872$\pm$0.049  & 0.021 & 1.203$\pm$0.088   & 0.300$\pm$0.024   & 1.45E-05  \\
BPR      & 9.021$\pm$0.068   & 2.744$\pm$0.045   & 0.036  & 1.820$\pm$0.061   & 0.609$\pm$0.027   & 6.95E-05  \\
RippleNet     & 9.294$\pm$0.027  & 2.401$\pm$0.036    & 0.040  & 1.882$\pm$0.041   & 0.624$\pm$0.025    & 8.07E-05 \\ \hline
PGPR     & 14.559$\pm$0.051  & 5.513$\pm$0.042    & -  & 7.003$\pm$0.032   & 2.843$\pm$0.030   & -  \\
Actor-Critic    & 14.821$\pm$0.043  & 5.730$\pm$0.051    & 0.912 & 6.924$\pm$0.044   & 2.796$\pm$0.031   & 0.949  \\
ADAC    & 15.856$\pm$0.053  & 5.863$\pm$0.048  & 0.718  & 7.501$\pm$0.022   & 3.010$\pm$0.058 & 0.748  \\ \hline
\end{tabular}
\end{table*}

\textbf{Data collection methodology.} 
There are a growing number of studies of RL-based recommender systems. To search relevant articles for the analysis, we adopted the following collection rules to include or exclude papers.
\renewcommand{\thefootnote}{\arabic{footnote}}
\begin{itemize}
\item Search terms: Our survey involves two keywords: Reinforcement Learning\footnote{Note that this survey focuses on recommender systems using full RL. Therefore, we did not include bandits, which are diferent from full RL.} and Recommender Systems. Accordingly, we mainly adopted the following search terms: 'Reinforcement Learning' AND 'Recommender Systems'. To find more research papers, we also used related RL algorithms as the search terms (\eg 'Q-learning', 'Policy Gradient', and 'Actor-Critic') instead of 'Reinforcement Learning', along with 'Recommender Systems'. Similarly, we used the search term 'Recommendation' instead of 'Recommender Systems', along with 'Reinforcement Learning'.

\item Search sources: We first used Google Scholar to find research papers with these search terms. We then increased the collection of relevant articles by the following academic databases: Science Direct, ACM Portal, Springer Link, and IEEE Xplore. Finally, we selected 98 related papers to include in this survey.

\item Publication type: Only high-level publications on RL-based recommendation from the international conferences and top journals were included in our survey.
\end{itemize}

\begin{figure}
  \centering
  \includegraphics[width=0.8\linewidth]{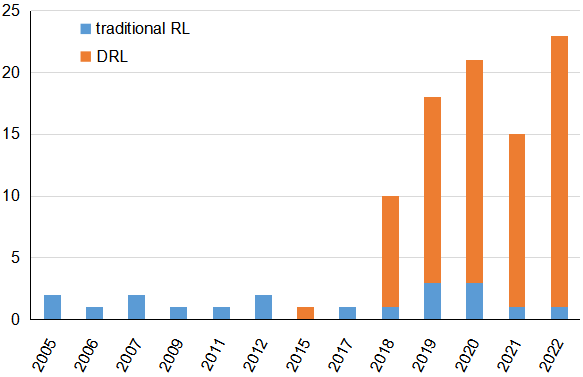}
  \caption{Distribution of year-wise publications (until October 2022) about traditional RL- and DRL-based recommendation methods.}
  \label{DP}
\end{figure}

The distribution of collected research papers over the years is shown in Fig.~\ref{DP}. There are a few publications on traditional RL-based recommendation methods from 2005 to 2017, with an increase in number of papers on DRL-based recommendation methods since 2018. The main reason is that DRL algorithms have proved to be the satisfactory solutions to some recommendation issues, and they have attracted much attention from the research community.

\textbf{Related work.} To facilitate the research about RL-based recommender systems, \cite{Afsar114} provides a review of the RL- and DRL-based algorithms developed for recommendations, and presents several research directions in top-$K$ recommendation, application architecture, and evaluation. Besides, \cite{Chen176} provides an overview of DRL-based recommender systems mainly according to model-based and model-free algorithms, and discusses the benefits and drawbacks of DRL-based recommendations. Nevertheless, it is necessary to make a more comprehensive overview and analysis of (D)RL-based recommender systems.

\textbf{Our contribution.} Different from \cite{Afsar114} and \cite{Chen176}, the main contributions made in this work are as follows.

\subsubsection{Comprehensive review} We summarize existing (D)RL algorithms applied in four typical recommendation scenarios, \ie interactive recommendation, conversational recommendation, sequential recommendation, and explainable recommendation. It could be helpful for readers to understand how (D)RL algorithms are applied in different recommendation systems. Moreover, from the RL perspective, the comprehensive survey of RL-based recommender systems follows three classes of RL algorithms: value-function, policy search, and Actor-Critic. This taxonomy of the literature is made based on the fact that these three types of methods have been widely applied in existing RL-based recommender systems. 

\subsubsection{Systematical analysis} We systematically analyze the challenges of applying (D)RL in recommender systems and relevant solutions, including environment construction (\eg the state representation and negative sampling), prior knowledge, reward function definition, learning bias (\eg the data or policy bias), and task structuring (\ie the task of RL can be decomposed into basic components or a sequence of subtasks). 

\subsubsection{Open directions} 
To facilitate future progress in this field, we put forward open issues of RL, analyze practical challenges of this field, and suggest possible future directions for the research and application of recommender systems.The open issues and emerging topics contain sampling efficiency, reproducibility, generalization, evaluation, biases, interpretability, safety and privacy.





The remainder of this paper is organized as follows. Section~\ref{RL} introduces the background of RL, defines related concepts, and lists commonly used approaches. Section~\ref{RLRS} presents a standard definition of the RL-based recommendation methods. Section~\ref{RS} provides a comprehensive review of the RL algorithms developed for recommender systems. Then, Section~\ref{Challenges} discusses the challenges and relevant solutions of applying RL in recommender systems. Next, Section~\ref{Discussion} discusses various limitations and potential research directions of RL-based recommender systems. Finally, Section~\ref{Conclusion} concludes this study.

\section{Overview of Reinforcement Learning}
\label{RL}
Different from supervised and unsupervised learning, RL \cite{Kiumarsi196} focuses on goal-directed learning that maximizes the total reward achieved by an agent when interacting with its environment. Trial-and-error and delayed rewards are two most important characteristics that distinguish RL from the other types of machine learning methods.

\begin{figure*}
  \centering
  \includegraphics[width=0.8\linewidth]{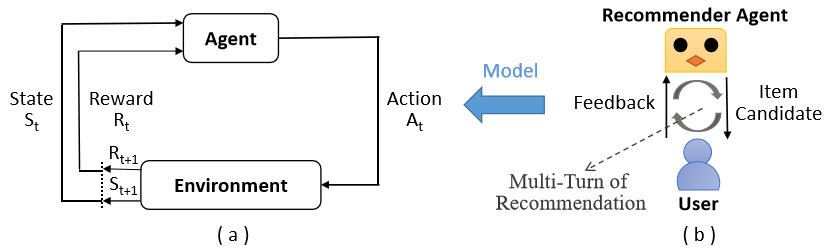}
  \caption{(a) Markov decision process formulated as the interaction between an agent and its environment \cite{Sutton72}. (b) RL-based recommender system models the interactive recommendation task as a Markov decision process.}
  \label{MDP}
\end{figure*}

In RL, the agent learns the optimal policy from its interactions with the environment to maximize the total reward for sequential decision making. As shown in Fig.~\ref{MDP}(a), the learning process of RL can be formulated as a Markov Decision Process (MDP)~\cite{Sutton72}. Formally, the MDP is defined as a 5-tuple $ <\mathcal{S},\mathcal{A},\mathcal{P},\mathcal{R},\gamma>$, where $ \mathcal{S} $ denotes a finite set of states, $ \mathcal{A} $ denotes a finite set of actions, $ \mathcal{P} $ is the state transition probabilities, $ \mathcal{R} $ denotes the reward function, and $ \gamma \in [0, 1]$ is a discount factor of the reward. 
At each time step $ t $, the agent receives an environment state $S_{t} \in \mathcal{S} $ and selects the corresponding action $ A_{t} \in \mathcal{A} $. Then, the agent receives a numerical reward $ R_{t+1} \in \mathcal{R} $ and makes itself into a new state $ S_{t+1} $.
Thus, the MDP forms a sequence $\tau $ as follows.
\begin{equation}\label{equ:1}
   \tau = \{S_{0}, A_{0}, R_{1}, S_{1}, A_{1}, R_{2}, \cdots, S_{T}\},
\end{equation}
where $ T $ is the maximum time step in a finite MDP. 
To maximize the return in the long run, the agent tries to select actions so that the cumulative reward it receives over the future is maximized. In this case, we introduce the concept of discounting. In general, the agent selects an action $ A_{t} $ to maximize the discounted return $ G_{t} $ \cite{Sutton72}:
\begin{equation}\label{equ:3}
   G_{t} = R_{t+1}+\gamma R_{t+2}+\cdots = \sum_{i=t+1}^{T}\gamma^{i-t-1}R_{i}.
\end{equation}
The discount factor $ \gamma $ affects the return. If $ \gamma = 0 $, the agent only maximizes immediate rewards then it reduces the return in the long run. As $ \gamma $ approaches 1, the agent takes future rewards into account more strongly.

Based on whether to use models and planning for solving RL problems, RL algorithms can be classified into two main groups, \ie model-free algorithms and model-based algorithms. The model-free algorithms directly learn the policy without any model of the transition function, whereas the model-based algorithms employ a learned or pre-determined model to learn the policy. On the other hand, according to the way of action conducted by the agent, RL algorithms fall into the following major groups, \ie value-function methods, policy search methods, and Actor-Critic methods.

\subsection{Value-function Approaches}
Many traditional RL methods generally achieve a global optimum return by obtaining the maximal value in terms of the best action. These methods are called value-function approaches, which utilize the maximal value to learn the optimal policy indirectly. Intuitively, the maximal value is generated by the best action $ a^{*} $ following the optimal policy $ \pi^{*} $, that is, the state-value under an optimal policy equals the expected return for the best action from the state. This is the Bellman equation for the optimal state-value function $ v_{\pi}^{*}(s) $, or called the Bellman optimality equation defined as:

\begin{equation}\label{equ:4}
\begin{split}
   v_{\pi}^{*}(s)&=\max_{a}\mathbb{E}[R_{t+1}+\gamma v_{\pi}^{*}(S_{t+1})|S_{t}=s,A_{t}=a]\\
   &=\max_{a}\sum_{s',r}p(s',r|s,a)\big[r+\gamma v_{\pi}^{*}(s')\big],
\end{split}
\end{equation}
where $ \mathbb{E}[\cdot] $ denotes the expected value of a random variable given by following the optimal policy $ \pi^{*} $.

Similarly, the Bellman optimality equation for the action-value function $ q_{\pi}^{*}(s,a) $ is defined by
\begin{equation}\label{equ:5}
\begin{split}
   q_{\pi}^{*}(s,a)&=\mathbb{E}[R_{t+1}+\gamma \max_{a'}q_{\pi}^{*}(S_{t+1},a')|S_{t}=s,A_{t}=a]\\
   &=\sum_{s',r}p(s',r|s,a)\big[r+\gamma \max_{a'} q_{\pi}^{*}(s',a')\big].
\end{split}
\end{equation}

Different from the optimal state-value function $ v_{\pi}^{*}(s) $, $ q_{\pi}^{*}(s,a) $ explicitly reflects the effects of the best action, which is more prone to be adopted by many algorithms.

In general, the optimal value function is estimated by Dynamic Programming (DP), Monte Carlo methods or temporal-difference (TD) learning, such as 
Sarsa \cite{Sutton72}, Q-Learning \cite{Watkins78}, and Deep Q-Networks (DQN) \cite{Mnih79}. Compared to policy-search methods, the value-function approaches are often of vastly reduced variance. Nevertheless, they are not suitable for complex application scenarios due to the slow convergence.

\subsection{Policy Search Methods}
In contrast to value-function approaches, policy-search methods directly optimize the policies, which are parameterized by a set of policy parameters $ \theta_{t} $.
These policy parameters can be updated to maximize the expected return with either gradient-free or gradient-based optimization methods \cite{Kai113}. As one of the most popular RL algorithms, the gradient-based optimization method is competent to solve complex issues. To search for the optimal policies, the gradient-based optimization method learns the policy parameters with the gradient of some performance measure $ J(\theta) $. Formally, the updates approximate gradient ascent in $ J(\theta) $ by
\begin{equation}\label{equ:7}
   \theta_{t+1}=\theta_{t}+\alpha \nabla J(\theta_{t}),
\end{equation}
where $ \nabla J(\theta_{t}) $ denotes a stochastic estimate that approximates the gradient of $ J(\theta_{t}) $ in terms of $ \theta_{t} $, and $ \alpha $ is the step size that influences the learning rate \cite{Sutton72}. Existing policy gradient methods generally follow the gradient updating strategy in Eq.~\eqref{equ:7}.

The policy gradient methods provide an appropriate equation proportional to the policy gradient, which may need Monte-Carlo based method of sampling their expectation that approximates the equation. Thus, the REINFORCE algorithm \cite{Williams71} that adopts the Monte-Carlo policy gradient method can be established by
\begin{equation}\label{equ:8}
\begin{split}
   \nabla J(\theta)&\propto\sum_{s}\mu(s)\sum_{a}\nabla_{\theta}\pi(a|s,\theta)q_{\pi}(s,a)\\
   &\doteq\mathbb{E}_{\pi}\Big[\sum_{a}\nabla_{\theta}\pi(a|S_{t},\theta)q_{\pi}(S_{t},a)\Big],
\end{split}
\end{equation}
where the symbol $ \propto $ refers to “proportional to”, the distribution $ \mu(s) $ is the on-policy distribution under the policy $ \pi $, and the gradients are column vectors of partial derivatives with respect to the policy parameter $ \theta $.


Some advanced algorithms have been proposed to address the shortcomings of policy-search methods. For example, policy gradient methods with function approximation \cite{Sutton73} ensure the stability of the algorithms. By adjusting super-parameters artificially or adaptively, Proximal Policy Optimization (PPO) Algorithms \cite{Schulman75} and Trust Region Policy Optimization (TRPO) \cite{Schulman84} speed up the convergence of the algorithms. Moreover, Guided Policy Search (GPS) \cite{Levine83} utilizes the path optimization algorithm to guide the training process of the policy gradient method, and thereby improves its efficiency. 

\subsection{Actor-Critic Algorithms}

There are a set of algorithms that incorporate the advantages of value-function approaches and policy search methods. They attempt to estimate a value function, meanwhile adopt the policy gradient to search in the policy space. Actor-Critic \cite{Konda74} is one of the most representative algorithms. It combines the policy-based method (\ie the actor) with the value-based approach (\ie the critic) to learn the policy and value-function together. 
The actor trains the policy according to the value function of the critic's feedback, while the critic trains the value function and uses the TD method to update it in one-step. The one-step Actor-Critic algorithm replaces the full return with the one-step return as
\begin{equation}\label{equ:9}
  \theta_{t+1}\doteq \theta_{t}+\alpha \big(R_{t+1}+\gamma\hat{v}(S_{t+1},\mathbf{w})-\hat{v}(S_{t},\mathbf{w})\big)\frac{\nabla_{\theta}\pi(A_{t}|S_{t},\theta_{t})}{\pi(A_{t}|S_{t},\theta_{t})},
\end{equation}
where $ \mathbf{w} $ is the state-value weight vector learned by the Actor-Critic algorithm \cite{Sutton72}. In Eq.~\eqref{equ:9}, $ \hat{v}(S_{t},\mathbf{w}) $ is a learned state value function that is used as the baseline. 
In recent years, many improved Actor-Critic algorithms have been developed, such as Asynchronous Advantage Actor-Critic (A3C) \cite{Mnih76}, Soft Actor-Critic (SAC) \cite{Haarnoja81}, Deterministic Policy Gradient (DPG) \cite{Silver82} and its variation Deep Deterministic Policy Gradient (DDPG) \cite{Lillicrap227}. Actor-Critic algorithms may alleviate the problem of sampling efficiency by experience replay \cite{Adam123}. However, due to the coupling of value evaluation and policy updates, the stability of these algorithms is unsatisfactory.

\section{Formulation of Recommendation Problem}
\label{RLRS}

In a typical recommender system, suppose there are a set of users $U$ and a set of items $I$, with $R\in \mathbb{R}^{X\times Y}$ denotes the user-item interaction matrix, where $X$ and $Y$ denote the number of users and items, respectively. Let $r_{t}^{ui}$ denote the interaction behavior between user $u$ and item $i$ at time step $t$. The recommender system aims to generate a predicted score $\hat{r}_{t}^{ui}$, which describes the user's preference on the item $i$.

Generally, we can formulate the recommendation task as a finite MDP \cite{Xin24, Liu146}, as shown in Fig.~\ref{MDP}(b). At each time step, the recommender agent interacts with the environment (\ie the user and/or logged data) by recommending an item to the user in the current state. At the next time step, the recommender agent receives feedback from the environment and recommends a new item to the user in a new state. The user’s feedback may contain explicit feedback (\eg purchase and rating) or implicit feedback (\eg user's browsing record from logged data). The recommender agent aims at learning an optimal policy with the policy network to maximize the cumulative reward.
More precisely, the MDP in recommendation scenario is a 5-tuple $ <\mathcal{S},\mathcal{A},\mathcal{P},\mathcal{R},\gamma>$, which can be defined as follows.

\begin{itemize}
\item \textbf{States} $ \mathcal{S} $. The finite state space describes the environment states in the fixed length history trajectories, in which $S_{t}=\{i_{1}, i_{2}, \cdots, i_{t}\}$ is an observed state from the sequence of interacted items at time step $t$.
\renewcommand{\thefootnote}{\arabic{footnote}}
\item \textbf{Actions} $ \mathcal{A} $. The discrete action space contains the whole set of recommended candidate items\footnote{Note that the action space may include other kinds of actions in different recommendation scenarios, \eg the selection of query attributes in conversational recommender systems, and the outgoing edges of entities in KG-based explainable recommender systems.}. An action $A_{t}$ is to recommend an item $i$ at time step $t$. In logged data, the action can be taken from the user-item interactions.

\item \textbf{Transition probability} $ \mathcal{P} $. It is the state transition probability matrix. The state transits from $s$ to $s'$ according to the probability $p(s',r|s,a)$ after the recommender agent receives the user's feedback (\ie the reward $r$).

\item \textbf{Reward function} $ \mathcal{R} $. Once the recommender agent takes action $a$ at state $s$, it obtains the reward $r(s, a)$ in accordance with the user’s feedback.

\item \textbf{Discount factor} $ \gamma $. It is the discount-rate parameter for future rewards.
\end{itemize}

We assume in online RL-based recommendation environment, the recommender agent recommends an item $ i_{t} $ to a user $ u $, while the user provides a feedback $ f_{t} $ for the recommender agent at the $t$-th interaction. The recommender agent obtains the reward $ r({S_{t},A_{t}}) $ associated with the feedback $ f_{t} $, and recommends a new item $ i_{t+1} $ to the user $ u $ at the next interaction.
Given the observation on the multi-turn of interactions, the recommender system generates a recommendation list. The recommender agent aims to learn a target policy $ \pi_{\theta} $ to maximize the cumulative reward of the sampled sequence \cite{Xin24}:
\begin{equation}\label{equ:10}
  \max_{\pi_{\theta}}\mathbb{E}_{\tau\sim\pi_{\theta}}[R_{\tau}] ,
\end{equation}
where  $ \theta $ refers to policy parameters, and  $R_{\tau}=\sum_{t=0}^{\left|\tau\right|}\gamma^{t}r({S_{t},A_{t}})$ denotes the cumulative reward with respect to the sampled sequence $ \tau = \{S_{0}, A_{0}, \cdots, S_{T}\} $.

The recommender agent often suffers from the high cost to learn the target policy by interacting with real users online. An alternative is to employ offline learning, which learns a behavior policy $ \pi_{\delta} $ from the logged data.
We should solve the policy bias to learn an optimal policy $ \pi^{*} $ when using the offline learning, since there is a noticeable difference between the target policy $ \pi_{\theta} $ and the behavior policy $ \pi_{\delta} $.

\begin{table*}[t]
  \centering
  \caption{Overview of RL algorithms for different recommender scenarios.}
\begin{tabular}{lllllll}
\hline
Scenario                          & Model                             & RL Algorithm                    &           &     RL Environment                    & Evaluation Strategy               & Dataset \\ \hline
\multirow{15}{*}{\begin{tabular}[c]{@{}l@{}}Interactive \\ Recommendation\end{tabular}}   & UWR \cite{Taghipour148}                 & \multirow{7}{*}{Value-function}  & Q-Learning                     &     model-free       & Offline     &    N/A      \\
                                               & Multi-with RL \cite{Zhang25}                  &                                  & DP           &             model-based          & Offline                &    ACM     \\
                                               & ARA \cite{Mahmood218}                  &                                  & DP           &             model-based          & Online \& Offline                &    N/A     \\
                                                                                              & DRCGR \cite{Gao51}                  &                                  & DQN           &  model-free                    & Offline                &   N/A      \\
                                                & UDQN \cite{Lei47}                  &                                  & DQN         &                model-free        & Offline                 &     ML100K, ML1M, YMusic   \\
                                               & FeedRec \cite{Zou19}               &                                  & DQN          &                    model-free    & Offline                  &     JD   \\
                                               & PDQ \cite{Zou90}                   &                                  & Q-Learning         &     model-based                    & Offline                 &  Taobao, Retailrocket      \\
                                               & KGQR \cite{Zhou27}                 &                                  & DQN          &              model-free          & Offline                      &   Book-Crossing, ML20M \\ \cline{2-7} 
                                               & SL+RL \cite{Huang58}               & \multirow{3}{*}{Policy Search} & REINFORCE                &    model-free       & Offline                   &    ML100K, ML1M, Steam  \\
                                               & RCR \cite{Zhang37}                 &                                  & REINFORCE        &    model-free   & Online \& Offline               &  Yelp, UT-Zappos50K \\
                                               & TPGR \cite{Chen15}                 &                                  & REINFORCE       &                model-free     & Offline                   &  ML10M, Netflix    \\ \cline{2-7} 
                                               & FairRec \cite{Liu26}               & \multirow{5}{*}{Actor-Critic}    & DPG       &                 model-free          & Offline                  &    ML100K, Kiva   \\
                                               & Attacks\&Detection \cite{Cao40}    &                                  & AC         &   model-free                       & Offline                    &   Amazon \\
                                               & SDAC \cite{Xiao100}                &                                  & AC        &                  model-based        & Offline                   &   RecSys, Kaggle   \\
                                               & AAMRL \cite{Yu152}                &                                  & AC          &                model-free        & Online                  &    UT-Zappos50K   \\
                                               & DRR \cite{Liu02}                   &                                  & AC          &                   model-free     & Online \& Offline               & \begin{tabular}[c]{@{}l@{}}ML1M, Yahoo! Music, \\ ML100K, Jester \end{tabular}\\ \hline
\multirow{9}{*}{\begin{tabular}[c]{@{}l@{}}Conversational \\ Recommendation\end{tabular}}    & EMC,BTD, EHL \cite{Maria41}           & \multirow{6}{*}{Value-function}   & \begin{tabular}[c]{@{}l@{}}Monte Carlo, \\ TD learning \end{tabular}                &    model-free        & Online \& Offline                  &   \begin{tabular}[c]{@{}l@{}}TRAVEL, PC, \\ CAMERA, CAR \end{tabular}    \\
                                  & ISRA \cite{Mahmood219}                  &                                  & DP           &             model-based          & Online                &    N/A     \\
                                   & EGE \cite{Wu190}    &                                  & Q-learning                &       model-free            & Online                   &   Shoes, Fashion IQ Dress   \\
                                                                      & MemN2N \cite{Tsumita161}    &                                  & DQN                  &    model-free             & Offline               &     Personalized Dialog     \\
                                   & SCPR \cite{Lei159}    &                                  & DQN                     &      model-free       & Online \& Offline                       &  Yelp, LastFM \\
                                               & UNICORN \cite{Yang154} &                                  & DQN            &      model-free         & Online \& Offline                 &   Yelp, LastFM, Taobao     \\ \cline{2-7}
                                               & CRM \cite{Sun21}                 & \multirow{2}{*}{Policy Search}   & REINFORCE         &  model-free    & Online \& Offline                  &  Yelp      \\
                                   & EAR \cite{Lei160}    &                                  & REINFORCE                       &     model-free      & Online \& Offline               & Yelp, LastFM \\ \cline{2-7} 
                                               & CRSAL \cite{Ren158}                   & Actor-Critic                     & A3C          &          model-free               & Offline                       &  \begin{tabular}[c]{@{}l@{}}DSTC2, CamRest676, \\ MultiWOZ 2.1 \end{tabular} \\  & RelInCo \cite{Ali212}                &                                  & AC          &               model-free        & Offline                & OpenDialKG, REDIAL \\ \hline
\multirow{9}{*}{\begin{tabular}[c]{@{}l@{}}Sequential \\ Recommendation\end{tabular}}     & SQN \cite{Xin24}                   & \multirow{4}{*}{Value-function}  & Q-learning           &      model-free          & Offline                   &   RC15, RetailRocket   \\
                                               & DEERS \cite{Zhao05}                &                                  & DQN          &               model-free        & Online \& Offline                & JD \\
                                                & NRRS \cite{Hong222}                &                                  & \begin{tabular}[c]{@{}l@{}}Monte Carlo \\ tree search \end{tabular}          &               model-based        & Offline                & Million Song \\
                                               & RLradio \cite{Moling42}            &                                  & R-Learning          &      model-based          & Online \& Offline                & N/A \\ \cline{2-7} 
                                               & KERL \cite{Wang10}                 & \multirow{3}{*}{Policy Search}   & Truncated PG       & model-free   & Offline                   &   Amazon, LastFM   \\
                                               & HRL+NAIS \cite{Zhang29}            &                                  & REINFORCE           &       model-free          & Offline                     & XuetangX   \\
                                               & DARL \cite{Lin133}                 &                                  & REINFORCE          &         model-free       & Offline                 &     XuetangX   \\ \cline{2-7} 
                                     & SAC \cite{Xin24}                 & \multirow{3}{*}{Actor-Critic}    & AC             &       model-free               & Offline                 &  RC15, RetailRocket   \\
                                      & DeepChain \cite{Zhao223}                &                                  & AC          &               model-based        & Online \& Offline                & JD \\
                                               & SAR \cite{Antaris189}                 &                                  & AC           &    model-free                    & Offline               &   \begin{tabular}[c]{@{}l@{}}Steam, Electronics, \\ ML10M, Kindle \end{tabular}     \\ \hline
                                               
\multirow{7}{*}{\begin{tabular}[c]{@{}l@{}}Explainable \\ Recommendation\end{tabular}}    & MT Learning \cite{Lu14}            & \multirow{5}{*}{Policy Search}   & REINFORCE            &         model-free        & Offline                &    Amazon, Yelp     \\
                                               & RL-Explanation \cite{Wang07} &                                  & REINFORCE          &       model-free           & Offline            &    Amazon, Yelp      \\
                                               & MKRLN \cite{Tao167}                &         & REINFORCE               &          model-free         & Offline            &  movie, book, KKBOX  \\                      & SAKG+SAPL \cite{Park178}                &         & REINFORCE                 &           model-free    & Offline          &  Amazon \\
                                            & PGPR \cite{Xian06}                &         & REINFORCE               &     model-free             & Offline 
                                                       & Amazon   \\\cline{2-7} 
                                               & ADAC \cite{Zhao57}                 & \multirow{2}{*}{Actor-Critic}    & AC          &             model-free             & Offline                &  Amazon     \\
                                               & AnchorKG \cite{Liu168}                 &                                  & AC          &                  model-free        & Offline              &    MIND, Bing News     \\ \hline
\end{tabular}
\label{TRS}
\end{table*}

\section{Recommender Systems by Reinforcement Learning}
\label{RS}

In this section, we summarize specific RL algorithms applied in four typical recommendation scenarios (\ie interactive recommendation, conversational recommendation, sequential recommendation, and explainable recommendation) following value-function, policy search, and Actor-Critic, respectively. An overview of related literature is shown in Table~\ref{TRS}. Note that some models/frameworks have an offline evaluation strategy in the offline experimental environment, whereas an online evaluation strategy denotes the experiments conducted on online communities or real users.

\subsection{Interactive Recommendation}
In a typical interactive recommendation scenario, a user $u$ is 
recommended an item $ i_{t} $ and provides a feedback $ f_{t} $ at the $t$-th interaction. The recommendation system recommends a new item $ i_{t+1} $ based on the feedback $ f_{t} $. Such an interactive process can be easily formulated as an MDP, where the recommender agent constantly interacts with the user and learns the policy from the feedback to improve the quality of recommendations \cite{Taghipour148}, as shown in Fig.~\ref{MDP} (b). Due to their nature of learning from dynamic interactions, RL algorithms have been widely adopted to solve interactive recommendation problems \cite{Taghipour148}.

\subsubsection{Value-function Approaches}

There is a common challenge of sample efficiency for RL algorithms, which may lead to inefficient learning of the recommendation policy. To address the limitation, Knowledge Graph (KG) is incorporated within RL algorithms for the interactive recommendation \cite{Zhou27}. KG can provide rich supplementary information, which reduces the sparsity of the user feedback. To this aim, KG enhanced Q-learning model is proposed to make the sequential decision efficiently. Besides, a DQN approach with double-Q \cite{Hasselt94} and dueling-Q \cite{Wang95} models the long-term user satisfaction. From the interaction history $o_{t}$ with item $i_{t}$ at time step $t$, the recommender agent obtains the reward $R_{t}$ and then stores the experience in the replay buffer $D$. The goal is to improve the performance of the Q-network by minimizing the mean-square loss function as follows.
\begin{equation}\label{equ:11}
   L(\theta_{q})= \mathbb{E}_{(o_{t}, i_{t}, R_{t}, o_{t+1})\sim D}[(y_{t}-q(S_{t}, i_{t};\theta_{q}))^{2}],
\end{equation}
where $ (o_{t}, i_{t}, R_{t}, o_{t+1}) $ refers to the learnt experience, and $ y_{t} $ denotes the target value of the optimal $ q^{*} $.

Moreover, \cite{Gao51} proposes a DQN-based recommendation framework that incorporates Convolutional Neural Network (CNN) and Generative Adversarial Networks (GAN) \cite{Ian101}, called DRCGR. It automatically learns the optimal policy based on both the positive and negative user feedback.

To optimize instant and long-term user engagement in recommender systems, \cite{Zou19} designs a novel Q-network that has three layers named raw behavior embedding layer, hierarchical behavior layer, and Q-value layer.
Moreover, Pseudo Dyna-Q (PDQ) \cite{Zou90} is proposed to ensure the stability of convergence and low computation cost of existing algorithms. The PDQ framework consists of two major components: a world model imitates user's feedback from the historical logged data and generates pseudo experiences. A recommendation policy based on Q-learning maximizes the expected reward by the pseudo experiences from the world model and logged experiences. 
Besides, a few attempts use the DP method to optimize the recommendation policies\cite{Mahmood218}. For example, \cite{Zhang25} adopts the value iteration algorithm to learn the true status value in a collaboration network.

The multi-step problem in interactive recommender systems has been studied in \cite{Lei47}, where the multi-step interactive recommendation is cast as a multi-MDP task for all target users. 
To model user-specific preferences explicitly, a biased User-specific Deep Q-learning Network (UDQN) is proposed by adding a bias parameter to capture the difference in the Q-values of different target users.

\subsubsection{Policy Search Methods}

Existing RL-based algorithms are often developed for short-term recommendation, whereas \cite{Huang58} employs DRL and Recurrent Neural Network (RNN) to improve the accuracy of long-term recommendation. Specifically, RNN is performed to simulate the sequential interactions between the environment (the user) and the recommender agent by evolving user states adaptively. This strategy can help tackle the cold-start issue in recommender systems. On the other hand, 
the interaction process is split into sub-episodes and reboot the accumulated reward of each sub-episode, which significantly improves the effectiveness of policy learning.

In text-based interactive recommender systems, user feedback with natural language usually causes undesired issues. For instance, the recommender system may violate user's preferences, since it ignores the previous interactions and thus recommends similar items. To handle these issues, a Reward Constrained Recommendation (RCR) model \cite{Zhang37} is proposed to incorporate user preferences sequentially. More specifically, the text-based interactive recommendation is formulated as a constraint-augmented RL problem, where the user feedback is taken as a constraint. To generalize from the constraints, a discriminator parameterized as the constraint function is developed to detect the violation of user's preferences. The policy is optimized by the policy gradient with baseline (\ie a general constraint), to learn constraints on violations of user's preferences. Based on the user's feedback on the recommended items, the recommender system utilizes constraints from these feedback to prevent undesired text generation.

Moreover, most existing RL methods fail to alleviate the issue of large discrete action space in interactive recommender systems, since there are a large number of items to be recommended. To solve this problem, \cite{Chen15} proposes a Tree-structured Policy Gradient Recommendation framework (TPGR) to achieve high effectiveness and efficiency for large-scale interactive recommendations. 
To maximize long-run cumulative rewards, the REINFORCE algorithm is utilized to learn the strategy for making recommendation decisions.

\subsubsection{Actor-Critic Algorithms}

Actor-Critic Algorithms have also been adopted extensively for interactive recommender systems in recent studies\cite{Liu02}. For instance, an RL-based framework (\ie FairRec) \cite{Liu26} is proposed to dynamically achieve a fairness-accuracy balance, in which the fairness status of the system and user's preferences combine to form the state representation. The FairRec framework contains two parts: an actor network and a critic network. The actor network generates the recommendation according to the fairness-aware state representation. The actor network is trained from the critic network, and updated by the DPG algorithm. Then, the critic network estimates the value of the actor network outputs. The critic network is updated by TD learning.


\cite{Cao40} addresses adversarial attacks in RL-based interactive recommender systems. They propose a general framework that consists of two models. In the adversarial attack model, the agent crafts adversarial examples following Actor-Critic or REINFORCE algorithm. In the encoder-classification detection model, the agent detects potential attacks based on the adversarial examples, employing a classifier designed by attention networks. Following \cite{Xian06}, the authors demonstrate the effectiveness of the proposed framework through judging whether the recommender system is attacked.

Online interactions in RL for interactive recommendation may hurt the user experiences. An alternative is to adopt logged feedbacks to perform offline learning. However, it usually suffers from some challenges, such as unknown logging policy and extrapolation error.
To deal with these challenges, \cite{Xiao100} proposes a stochastic Actor-Critic method based on a probabilistic formulation, and present some regularization approaches to reduce the extrapolation error.

In interactive recommender systems, utilizing multi-modal data can enrich user feedback. To this end, \cite{Yu152} proposes a vision-language recommendation approach that enables effective interactions with the user by providing natural language feedback. In addition, an attribute augmented RL is introduced to model explicit multi-modal matching. More specifically, the multi-modal data $ A_{t} $ (\ie the action of recommending items at time $ t $) and $ x_{t} $ (\ie natural language feedback from users at time $ t $) are leveraged in the proposed approach. Then a recommendation tracker, consisting of a feature extractor and a multi-modal history critic, is designed to enhance the grounding of natural language to visual items. The recommendation tracker may track the user's preferences based on a history of multi-modal matching rewards. The policy is updated via the Actor-Critic algorithm for recommending the items with desired attributes to the user.

\subsection{Conversational Recommendation}

\begin{figure*}
  \centering
  \includegraphics[width=0.9\linewidth]{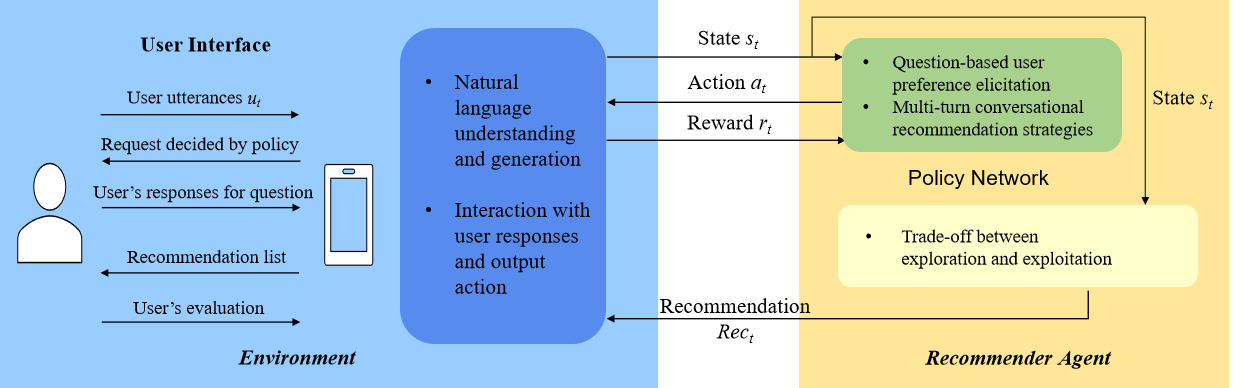}
  \caption{The common framework of conversational recommendation model based on RL. The user interface extracts user intention from the utterances during the conversation \cite{Gao87}. The policy network learns the optimal policy from dialogue states. The recommender system is trained with user intention, item features, and dialogue states, to make personalized recommendations \cite{Sun21, Ren158, Tsumita161}.}
  \label{FCR}
\end{figure*}

Contrary to interactive recommender systems where users receive information passively (\ie the recommendation system is dominant), conversational recommender systems interact with users actively. They explicitly acquire users' active feedback and make recommendations that users really like. To achieve this objective, different from interactive recommender systems that recommend items from each interaction, conversational recommender systems \cite{Hu208} usually recommend items after communicating with users by real-time multi-turn interactive conversations, based on natural language understanding and generation. In this case, there is a critical issue of trade-offs between exploration and exploitation for conversation and recommendation strategies \cite{Gao87}. Conversational recommender systems explore the items unseen by a user to capture the user's preferences by multi-turn interactions. However, compared to exploiting the related items that have already been captured, exploring the items that may be unrelated will harm the user experience. RL provides potential solutions to address this challenge. As shown in Fig.~\ref{FCR}, the policy network stimulates the reward centers in conversational recommender systems, integrating exploration with exploitation in multi-turn interactions.

\subsubsection{Value-function Approaches}

At each conversation turn of conversational recommendation, the policy learning usually aims to decide what to ask, which to recommend, and when to ask or recommend. \cite{Yang154} introduces the UNICORN (\ie UNIfied COnversational RecommeNder) model that employs an adaptive DQN method to cast these decision-making processes as a unified policy learning task. 
To adapt the conversational strategy, \cite{Mahmood219} adopts the DP method to yield better policies that assist user behaviors more efficiently. Besides, \cite{Tsumita161} proposes an RL-based dialogue strategy that utilizes recommendation results based on the user's utterances, whose intention is estimated by a Long Short-Term Memory (LSTM) network.

It is interesting that conversational recommender systems can use incremental critiquing as a powerful type of user feedback, to retrieve the items in line with the user's soft preferences at each turn. From this point of view, it needs a related quality measure for the recommendation efficiency. To achieve this objective, a novel approach is proposed to improve the quality measure by combining a compatibility score with similarity score \cite{Maria41}. To evaluate the compatibility of user critiques, exponential reward functions are presented by Monte Carlo and TD methods based on the user specialization. Moreover, a global weighting of user's preferences is proposed to enhance the critiquing quality, which brings about faster convergence of the similarity. By using the combination of these two scores, the conversational recommender system improves its robustness against the noisy user data.

However, it is difficult to capture users’ preferences over time since the recommender system usually only obtains partial observations of the users’ preferences. In this case, we can formalize the observed interactions as a Partially Observable Markov Decision Process (POMDP). Afterwards, the Estimator-Generator-Evaluator (EGE) \cite{Wu190} trains the POMDP with Q-learning to track users’ preferences and generates the next recommendations at each iteration.


The aforementioned conversational recommender systems usually utilize user feedback in implicit ways. Instead, to make full use of user preferred attributes on items in an explicit way, a Simple Conversational Path Reasoning (SCPR) framework \cite{Lei159} is proposed to conduct interactive path reasoning for conversational recommendation on a graph. The SCPR obtains user preferred attributes more easily, by pruning off irrelevant candidate attributes following user feedback based on a policy network. The policy network inputs the state vector $s$ and outputs the action-value $ Q(s, a)$, referring to the estimated reward for asking action $a_{ask}$ or recommending action $a_{rec}$. The policy is optimized by the standard DQN to achieve its convergence. Different from EAR \cite{Lei160}, SCPR uses the adjacent attribute constraint on the graph to reduce the search space of attributes, such that the decision-making efficiency can be improved.

\subsubsection{Policy Search Methods}

To model the user's current intention and long term preferences for personalized recommendations, \cite{Sun21} utilizes deep learning technologies to train a conversational recommender system. Specifically, a deep belief tracker extracts a user intention by analyzing the user's current utterance. Meanwhile, a deep policy network guides the dialogue management by the user's current utterance and long-term preferences learned by a recommender module.
The framework of the proposed Conversational Recommender Model (CRM) is composed of three modules: a natural language understanding with the belief tracker, dialogue management based on the policy network, and a recommender designed based on factorization machine \cite{Rendle21}. More specifically, at first, the belief tracker converts the utterance (\ie the input vector $ z_{t} $) into a learned vector representation (\ie $ S_{t} $) by LSTM network. Afterwards, to maximize the long-term return, they use the deep policy network to select a reasonable action from the dialogue state at each turn. The REINFORCE algorithm is adopted to optimize the policy parameter. Finally, the factorization machine is utilized to train the recommendation module, which generates a list of personalized items for the corresponding user.

Distinct from previous studies that ignore how to adapt the recommended items for user feedback, \cite{Lei160}
proposes a unified framework called Estimation Action Reflection (EAR), in which a Conversation Component (CC) intensively interacts with a Recommender Component (RC) in the three–stage process. Specifically, the framework starts from the estimation stage, the RC guides the action of CC by ranking candidate items. Afterwards, at the action stage, the CC decides which questions to ask in terms of item features and makes a recommendation. The conversation action is performed by a policy network, which is optimized via the REINFORCE algorithm. When a user rejects the recommended items that are made at the action stage, the framework moves to the reflection stage for adjusting its estimations. 
Extensive experiments demonstrate the proposed EAR outperforms CRM \cite{Sun21} according to not only fewer conversation turns but also better recommendations, mainly because the candidate items adapt to user feedback in the interaction between the RC and CC.

\subsubsection{Actor-Critic Algorithms}

To address the training issue of task-oriented dialogue systems, a Conversational Recommender System with Adversarial Learning (CRSAL) \cite{Ren158} is proposed to fully enable two-way communications. 
The process of CRSAL is divided into three stages. In the information extraction stage, a dialogue state tracker firstly infers the current dialogue belief state $b_{t}$ from the user's utterances. Afterwards, a neural intent encoder extracts and encodes the user’s utterance intention $z_{t}$. Finally, a neural recommender network derives recommendations from item features and dialogue states.
In the conversational response generation stage, a neural policy agent (\ie the actor) generates human-like action in the current dialogue state, and a natural language generator updates conversational responses from the critic network by the selected action. In the RL stage, the decision procedure of dialogue actions can be formulated as a Partially Observable MDP (POMDP). The agent selects the best action in each conversation round under the long-term policy. To this end, an adversarial learning mechanism based on the A3C algorithm is developed to fine-tune the actions generated by the agent, which employs the discriminator to train the optimal parameters of the proposed model.

In conversational recommender systems, valuable information from user's utterances is often conducive to the retrieval performance. For example, \cite{Ali212} proposes an RL-based model to extract relevant information from the context of the conversation. The model introduces two Actors: a selector-Actor finds the most relevant words for the target of the conversation, and an arrangement-Actor returns the related order of words based on the user's utterances. Both Actors are trained by the Actor-Critic algorithm.

\subsection{Sequential Recommendation}

Unlike interactive recommendation methods that generate recommendations based on the user's feedback via constant interactions, sequential recommender systems predict the user's future preference and recommend the next item given a sequence of historical interactions. Let $i_{i:n}^u = i_{1}^u \rightarrow i_{2}^u \rightarrow \cdots \rightarrow i_{n}^u $ denote the user-item interaction sequence, where $n$ is the sequence length. The sequential recommender system aims to recommend the next item $\hat{i}$ that is not in the historical interactions.
Some attempts have revealed that RL algorithms deal well with the sequential recommendation problems, since such problems can be naturally formulated as an MDP to predict the user's long-term preferences. In this case, the recommender agent easily performs a sequence of ranking, which usually learns the optimal policy from the logged data with off-policy methods.

\subsubsection{Value-function Approaches}

In sequential recommendation, it is essential to fuse long-term user engagement and user-item interactions (\eg clicks and purchases) into the recommendation model training. Learning the recommendation policy from logged implicit feedback by RL is a promising direction. However, there exists challenges in implementation due to a lack of negative feedback and the pure off-policy setting. \cite{Xin24} presents a novel self-supervised RL method to address the challenges.
More precisely, the next item recommendation problem is modeled as an MDP, where the recommender agent sequentially recommends items to related users to maximize the cumulative reward.
To optimize the recommendation model, they define a flexible reward function that contains purchase interactions, long-term user engagement, and item novelty. 
Based on this method, the authors develop a self-supervised Q-learning model to train two layers with the logged implicit feedback. Similarly, \cite{Moling42} leverages R-Learning \cite{Schwartz144} to develop a music recommender system named RLradio. It exploits both explicitly revealed channel preferences and user's implicit feedback, \ie the user actually listens to a music track that played in the recommended channel. \cite{Hong222} also leverages the wireless sensing and RL algorithm to improve the user experience in music recommender systems, in which user's preferences are explored by the Monte Carlo tree search method. 

Moreover, in \cite{Zhao05}, a novel DQN is built for the proposed framework, where Gated Recurrent Unit (GRU) captures the user's sequential behaviors as positive state $ s_{+} $, and negative state $ s_{-} $ is obtained by a similar way. The positive and negative signals are fed into the input layer separately, which assists the new DQN to distinguish contributions of the positive and negative feedback in recommendations.



\subsubsection{Policy Search Methods}
It is a non-trivial problem to capture the user's long-term preferences in sequential recommender systems, since the user-item interactions may be sparse or limited. Thus, it is unreliable for RL algorithms to learn user interests by using random exploration strategies. To overcome these issues, \cite{Wang10} proposes a Knowledge-guidEd Reinforcement Learning (KERL) framework that adopts an RL model to make recommendations over KG.
In the MDP modeled by KERL, the environment contains the information of interaction data and KG, which are useful for the sequential recommendation. In this case, the agent selects an action $ a $ in state $ S_{t} $ for recommending an item $ i_{t+1} $ to a user $ u $. During each recommendation process, the agent obtains an intermediate reward, which is defined by integrating sequence-level and knowledge-level rewards.

Moreover, \cite{Zhang29} analyzes users' sequential learning behaviors and points out that the attention-based recommender systems perform poorly when the users enroll in diverse historical courses, because the effects of the contributing items are diluted by many different items. To remove noisy items and recommend the most relevant items at the next time, a profile reviser with two-level MDPs is designed. In this profile reviser, a high-level task decides whether to revise the user profile, and a low-level task decides which item should be removed. The agent in the proposed Hierarchical Reinforcement Learning (HRL) framework performs hierarchical tasks under the revising policy, which is updated by the REINFORCE algorithm. 
In addition, for capturing users’ dynamic preferences in sequential learning behaviors, \cite{Lin133} proposes a dynamic attention mechanism, which combines with the HRL-based profile reviser to distinguish the effects of contributing courses in each interaction. As a result, the proposed model achieves more accurate recommendations than HRL \cite{Zhang29}.

\subsubsection{Actor-Critic Algorithms}

As mentioned before, \cite{Xin24} proposes a novel self-supervised RL method to learn the recommendation policy from logged implicit feedback in sequential recommendation. To optimize the recommendation model, a Self-supervised Actor-Critic (SAC) model treats the self-supervised layer as an actor to perform ranking and takes the RL layer as a critic to estimate the state-action value.

However, both SQN and SAC \cite{Xin24} utilize a fixed length of interaction sequences as input to train the models, which affects the recommendation accuracy, because users usually have various sequential patterns. To address this issue, \cite{Antaris189} proposes a Sequence Adaptation model via deep Reinforcement learning (SAR) to adjust the length of interaction sequences. In particular, the RL agent performs the selection of a sequence length (\ie action) in a personalized manner in the actor network. Finally, a joint loss function is optimized to align the cumulative rewards of the critic network with the recommendation accuracy.

\begin{figure*}
  \centering
  \includegraphics[width=0.9\linewidth]{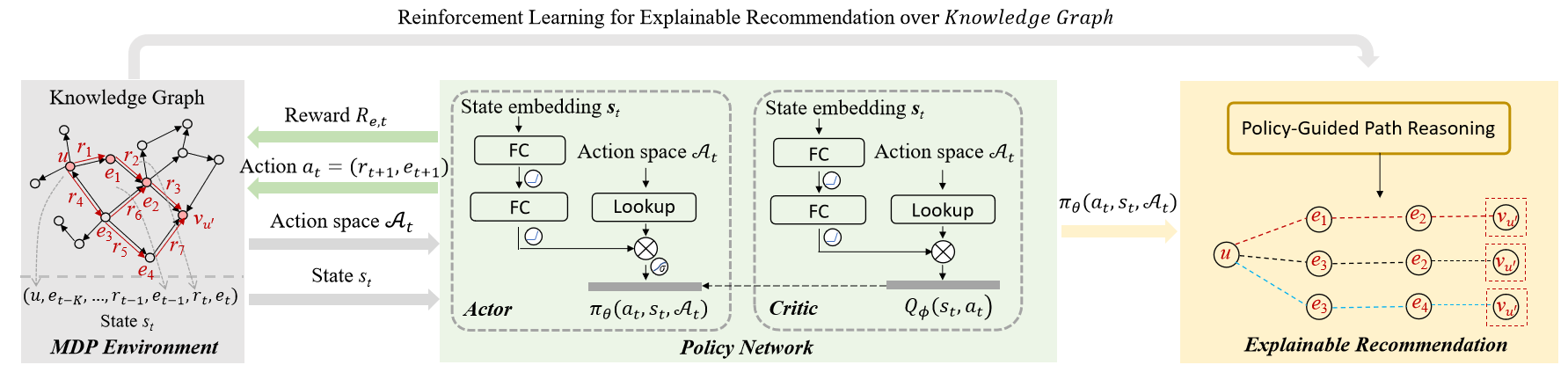}
  \caption{The RL algorithms with KG for explainable recommendation \cite{Xian06, Zhao57}. The models aim to learn a path-finding policy over KG. The learned policy is adopted for the interpretable reasoning paths that make accurate recommendations to the given user.}
  \label{KGRL}
\end{figure*}

\subsection{Explainable Recommendation}

The objective of explainable recommendation is to solve the problem of interpretability in the recommender systems. The explainable recommender systems not only provide high-quality recommendations but also generate relevant explanations for recommendation results \cite{Wang210}.
In particular, the visual explanation seems more intelligent when a recommender system conducts path reasoning over KG, because KG contains rich relationships between users and items for intuitive explanations. Fig.~\ref{KGRL} illustrates the common framework of RL with KG for the explainable recommendation. The KG is cast as a part of the MDP environment, and the agent performs path reasoning for recommendations \cite{Xian06, Zhao57}.

In the survey on explainable recommendation \cite{Zhang86}, the explainable recommendation methods are divided into five classes: explanation with relevant items or users, feature-based explanation, social explanation, textual sentence explanation, and visual explanation. The emphasis of this section is to survey RL applied to the explainable recommendation, where the approaches focus on textual sentence explanation and visual explanation.

\subsubsection{Policy Search Methods}
In terms of textual sentence explanation, \cite{Lu14} proposes a multi-task learning framework that simultaneously makes rating predictions and generates explanations from users' reviews. The rating prediction is performed by a context-aware matrix factorization model, which learns latent factor vectors of users and items. The recommendation explanation is employed by an adversarial sequence-to-sequence model, in which GRU generates personalized reviews from observed review documents, and adversarial training for the review generation is optimized by the REINFORCE algorithm.

To control the explanation quality flexibly, \cite{Wang07} designs an RL framework to generate sentence explanations. In the proposed framework, there are couple of agents instantiated with attention-based neural networks. The task of one agent is to generate explanations, while the other agent is responsible for predicting the recommendation ratings based on the explanations. The environment mainly consists of users, items, and prior knowledge. In this work, the recommendation model is treated as a black box. The agents extract the interpretable components from the environment to generate effective explanations.

Furthermore, to leverage the image information of entities, rather than focusing on rich semantic relationships in a heterogeneous KG,  \cite{Tao167} presents a Multi-modal Knowledge-aware RL Network (MKRLN), where the representation of recommended path consists of both structural and visual information of entities. The recommender agent starts from a user and searches suitable items along hierarchical attention-paths over the multi-modal KG. These attention-paths can improve the recommendation accuracy and explicitly explain the reason for recommendations.

The real-world KG is generally enormous, with a focus on finding more reasonable paths in the graph. From this aspect, \cite{Xian06} proposes a reinforcement KG reasoning approach called Policy-Guided Path Reasoning (PGPR), which performs recommendations and explanations by providing actual paths with the REINFORCE algorithm over the KG. Specifically, the recommendation problem is treated as a deterministic MDP over the KG. In the training stage, following a soft reward and user-conditional action pruning strategy, the agent starts from a given user, and learns to reach the correct items of interest. In the inference stage, the agent samples diverse reasoning paths for recommendation by a policy-guided search algorithm, and generates genuine explanations to answer why the items are recommended to the user.

Distinct from previous explainable recommendation methods that use KGs, \cite{Park178} focuses on sentiment on relations in the KG. To obtain more convincing explanations with sentiment analysis, a Sentiment-Aware Knowledge Graph (SAKG) is constructed by analyzing users' reviews and ratings on items. Moreover, a Sentiment-Aware Policy Learning (SAPL) method is introduced to make recommendations and guide the reasoning over the SAKG. Experimental results demonstrate that the proposed framework outperforms state-of-the-art baselines (\eg PGPR \cite{Xian06}), in terms of both accuracy and explainability.

\subsubsection{Actor-Critic Algorithms}

Differing from PGPR \cite{Xian06}, the ADversarial Actor-Critic (ADAC) model \cite{Zhao57} is proposed to identify interpretable reasoning paths.
By learning the path-finding policy, the actor obtains its search states over the KG and potential actions. The actor obtains the reward $ R_{e,t} $ if the path-finding policy from the current state fits the observed interactions.
To integrate the expert path demonstrations, they design an adversarial imitation learning module based on two discriminators (\ie meta-path and path discriminators), which are trained to distinguish the expert paths from the paths generated by the actor. When its paths are similar to the expert paths in the meta-path or path discriminator, the actor obtains the reward ($ R_{m,t} $ or $ R_{p,t} $) from the imitation learning module.
The critic merges these three rewards (\ie $ R_{e,t} $, $ R_{m,t} $, and $ R_{p,t} $) to estimate each action-value accurately. Later, the actor is trained with an unbiased estimate of the reward gradient through the learned action-values. Finally, the major modules of ADAC are jointly optimized to find the demonstration-guided path, which brings accurate recommendations.

Although existing works (\eg PGPR \cite{Xian06} and ADAC \cite{Zhao57}) combine KG and RL to enhance recommendation reasoning, they are not suitable for news recommendation, where a news article usually contains multiple entities. To address this challenge, a recommendation reasoning paradigm, named AnchorKG, is proposed to employ anchor KG path to make news recommendation \cite{Liu168}. The AnchorKG model consists of two parts. An anchor graph generator captures the latent knowledge information of the article, which leverages $k$-hop neighbors of article entities to learn high-quality reasoning paths. On the other hand, an AC-based framework is developed to train the anchor graph generator. Finally, the model performs a multi-task learning process to optimize both the anchor graph generator and the recommendation task jointly.

\subsection{Discussion on RL Aspect}
As mentioned above, the comprehensive survey of RL-based recommender systems follows three classes of RL algorithms: value-function, policy search, and Actor-Critic. From the RL perspective, we make a comparison of these three types of methods when they are applied in the recommender systems.

\begin{itemize}
\item \textbf{Value-function approaches} depend heavily on the samples to learn the optimal policy. Thus, these approaches are suitable for small discrete action spaces and are often applied in small-scale recommender systems, such as traditional interactive recommendation and conversational recommendation. However, real-world recommender systems usually contain large action spaces, which leads to the slow convergence of RL when they utilize value-function approaches to make recommendations.

\item \textbf{Policy search methods} directly optimize the policy without relying on the value function. They are only suitable for continuous action spaces due to the policy gradient, which results in high variances. However, benefitting from fast convergence properties, policy search methods are competent in large-scale recommender systems, such as sequential recommendation and explainable recommendation, especially KG-based recommendation.

\item \textbf{Actor-Critic algorithms} incorporate the advantages of value-function approaches and policy search methods. Nevertheless, Actor-Critic algorithms usually cause potential information loss since they map discrete action spaces into continuous action spaces by the policy network \cite{Chen176}, which ensures the policy is differentiable with respect to its parameters. Hence, Actor-Critic algorithms are rarely applied in the recommender systems that focus on the convincing recommendations, \eg conversational recommendation and explainable recommendation.
\end{itemize}

RL-based recommendation models can also be divided into model-based and model-free algorithms. As shown in Table~\ref{TRS}, a few recommendation models adopt model-based algorithms, while most existing recommendation models utilize model-free algorithms.

\begin{itemize}
\item \textbf{Model-based algorithms} (\eg DP and heuristic search) require a model to represent the environment of the recommender system, thus the agent relies on planning and guarantees sample efficiency. Nevertheless, such algorithms often result in biased estimations since the environment of recommender systems dynamically changes. Moreover, the transition probability is deterministic in model-based algorithms. Therefore, these algorithms are not applicable to real-world recommender systems.

\item \textbf{Model-free algorithms} (\eg TD, DQN, and REINFORCE) generally achieve better recommendation performances, because the agent mainly relies on learning from previous experiences. The drawback of such algorithms is sample inefficiency, \ie they require a large number of samples to ensure the convergence of the algorithms.
\end{itemize}

\section{Challenges in Reinforcement Learning based Recommendation Approaches}
\label{Challenges}

As noted above, RL aims to maximize long-run cumulative rewards by learning the optimal policy from the interactions between the agent and its environment. Thus, it relies not only on the environment but also on prior knowledge. In the recommendation applications, RL approaches often suffer from various challenges.
To this end, many researchers put forward relevant solutions to different issues. In the following, we summarize relevant studies from five aspects: environment construction, prior knowledge, reward function definition, learning bias, and task structuring.

\subsection{Environment Construction}
In RL, the agent observes states from the environment, then conducts relevant actions under the policy, and receives a reward from the environment. Applied to recommender systems, the policy training in the environment is often confronted with many unpredictable situations, due to the need for exploration. In this case, environment construction is critical to learn the optimal recommendation policy.

\subsubsection{State Representation}
The state representation plays an important role in RL to capture the dynamic information during the interactions between users and items, since the environment state is a key component of MDP. However, most current RL methods focus on policy learning to optimize recommendation performance. To effectively model the state representation, \cite{Liu02} proposes a DRL-based recommendation framework termed DRR, where four state representation schemes are designed to learn the recommendation policy (the actor network) and the value function (the critic network).
Specifically, the DRR-p scheme employs the element-wise product operator to learn the pairwise local dependency between items. DRR-u scheme incorporates the pairwise interactions between users and items into the DRR-p scheme. DRR-ave scheme concatenates the user embedding, the average pooling result of items, and the user-item interactions into a whole vector to describe the state representation. In the DRR-att scheme, a weighted average pooling is conducted by attention networks.
The actor network conducts a ranking action $ a=\pi_{\theta}(s) $ according to the state representation $ s $, and generates a corresponding $Q$-value by the action-value function $ Q_{\pi}(s, a) $. The critic network leverages a DQN parameterized as $ Q_{w}(s,a) $ to approximate $ Q_{\pi}(s, a) $. Based on DPG \cite{Silver82}, the actor network can be updated by the policy gradient via
\begin{equation}\label{equ:25}
   \nabla_{\theta} J(\pi_{\theta})\approx \frac{1}{N}\sum_{t}\nabla_{a}Q_{w}(s,a)|_{s=S_{t},a=\pi_{\theta}(S_{t})}\nabla_{\theta}\pi_{\theta}(s)|_{s=S_{t}},
\end{equation}
where $ J(\pi_{\theta}) $ denotes the expectation of all possible $Q$-values following the policy $ \pi_{\theta} $, and $ N $ is the batch size. 

Moreover, \cite{Shang16} introduces the DEMER (\ie DEconfounded Multi-agent Environment Reconstruction) framework, which assumes that the environment reconstruction from the historical data is powerful in RL-based recommender systems. DEMER randomly samples one trajectory from the observed data, and then forms its first state as the initial observation. It finally generates a policy-generated trajectory by the TRPO algorithm, considering the confounder embedded policy as a role of hidden confounders in the environment.

It may be practical to formulate a simulation environment of recommendation as an RL gym. To achieve this objective, PyRecGym \cite{Bichen147} is designed for RL-based recommender systems to support standard test datasets and common input types. In the PyRecGym, the states of the environment refer to user profiles, items, and interactive features with contextual information. The RL agent interacts with a gym engine in the current state, and obtains related feedback from the gym engine according to the reward. PyRecGym contains three major functions: initialization function that initializes the initial state of the environment for the user, reset function that resets the user state in the next episode and returns the initial state to the RL agent, and step-function that reacts to the agent’s action and returns the next state.

\subsubsection{Knowledge Graph Leverage}

Recent advances in KG have attracted increasing attention in RL-based recommender systems. The KG-enhanced interaction fashion enriches user-item relations by the structural knowledge to describe the MDP environment. Besides, the multi-hop paths in a KG contribute to the reasoning process for explainable recommendation.

The KERL framework \cite{Wang10} incorporates KG into RL-based recommender systems, with the aim of predicting future user's preferences and addressing the sparsity. 
To learn the user's preferences from sparse user feedback, \cite{Zhou27} proposes a KG-enhanced Q-learning model for interactive recommender systems. Instead of learning the policy by trial-and-error search, the model utilizes the knowledge of correlations among items learned from KG to enrich the state representation of both items and users. Thus, it spreads the user's preferences among the correlated items over KG.

\cite{Xian06} made the first attempt to leverage KG and RL for the explainable recommendation. They develop a unified framework (\ie PGPR) to provide actual recommendation paths guided by the policy network in a KG. 
To further improve the convergence of the PGPR approach, \cite{Zhao57} proposes a demonstration-based KG reasoning framework named ADAC, in which imperfect path demonstrations are extracted to guide path-finding. The ADAC model aims to identify interpretable reasoning paths for accurate recommendations.

Another interesting work is the negative sampling by KG Policy Network \cite{Wang18}. It is incorporated into the recommendation framework to explore informative negatives over KG. 
The recommender (\ie matrix factorization) and the sampler (\ie KG Policy Network) are jointly trained by the iteration optimization. The recommender parameters are updated by the stochastic gradient descent (SGD) method, and the sampler parameters are updated via the REINFORCE algorithm.

\subsubsection{Negative Sampling}

Most existing studies extract negative sampling from the unobserved data to assist the training of the recommendation model. However, they often fail to yield high-quality negative samples to reflect the user's needs, which provides an important clue on the environment construction for RL.
To discover informative negative feedback from the missing data, \cite{Wang18} proposes a KG policy network for knowledge-aware negative sampling, which employs an RL agent to search high-quality negative instances with multi-hop exploring paths over the KG. 
Similarly, to learn the sampling strategy from the missing negative signals, the supervised negative Q-learning method \cite{Xin206} trains the RL algorithm with a supervised sequential learning method to sample negative items.

Moreover, the user exposure data, which records the history interactions based on implicit feedback, is also beneficial to train negative samples. From this point of view, \cite{Ding35} introduces a recommender-sampler framework, where the sampler samples candidate negative items as the output, and the recommender is optimized by the SGD method to learn the pairwise ranking relation between a ground truth item and a generated negative item. After obtaining the multiple rewards, the sampler is optimized by the REINFORCE algorithm to generate both real and hard negative items. 
Moreover, \cite{Zhao200} uses a CF-based pre-training method to sample negative items for RL-based recommender systems.

\subsubsection{Social Relation}
Traditional recommender systems often suffer from two main challenges, \ie cold start and data sparsity. A promising way for alleviating these issues is to utilize users' social relationships to model users' preferences efficiently. Thus, the agent can sample users’ social relationships and deliver them to the environment.

Applied to RL-based recommendation, \cite{Lei48} integrates social networks into the estimation of action-values. More specifically, a social matrix factorization method is proposed to describe the high-level state/action representations. 
To learn more relevant hidden representations from personal preferences and social influence, an enhanced SADQN model is developed to utilize additional neural layers to summarize potential features from the hidden representations, and then predict the final action-value with the summarized features. 
Moreover, \cite{Lu50} proposes a Social Recommendation framework based on Reinforcement Learning (SRRL) to identify reliable social relationships for the target user. 
In particular, SRRL adaptively samples the social friends to improve the recommendation quality with user feedback, since the reward is always real-time.

\subsection{Prior Knowledge}

The combination of imitation learning with RL can be named apprenticeship learning \cite{Abbeel109}, which utilizes demonstrations to initialize the RL. In particular, the Inverse Reinforcement Learning (IRL) algorithm \cite{Andrew102} often assumes that the expert acts to maximize the reward, and derives the optimal policy from the learned reward function. In RL-based recommender systems, the reward signals are usually unknown, since users scarcely offer feedback. On the other hand, IRL algorithms are good at reconstructing the reward function for the optimal policy from users' observed trajectories.

To improve the novelty of the next-POI (\ie Point-of-Interests) recommendation that boosts users' interests, \cite{David55} adopts a novel IRL algorithm termed Maximum Log-likelihood (MLIRL) \cite{Babes110} to model the unknown user's preferences based on the state features. This method exploits the knowledge of the user's preferences to estimate an initial reward function that justifies the observed trajectories, and optimizes the user's behavior by a gradient ascent method.

To mine users' interests in online communities efficiently, \cite{Liang52} designs a reinforced user profiling for recommendations by employing both data-specific and expert knowledge, where the agent employs random search to find data-specific paths in the environment. These meta-paths contain expert knowledge and semantic meanings for searching useful nodes. 

Distinct from the top-$K$ recommendation with the target of ranking optimization, the novel exact-$K$ recommendation focuses on combinatorial optimization. It may be more suitable to address the recommendation problems in application scenarios. Towards this end, \cite{Gong28} designs an encoder-decoder framework named Graph Attention Networks (GAttN). The proposed framework learns the joint distribution of the $K$ items and outputs an optimal card that contains these $K$ items. To train the GAttN efficiently, an RL from demonstrations method integrates behavior cloning \cite{Torabi108} into RL.

\subsection{Reward Function Definition}

The agent behavior in RL indirectly relies on the reward function, while in practice, it is a key challenge to define a promising reward function. Beyond the need to learn the policy by intermediate rewards, the reward definition based on the specific demands in different recommendation scenarios is necessary. For instance, \cite{Xin24} defines a flexible reward function that contains purchase interactions, long-term user engagement, and item novelty for the relevant recommendation tasks. \cite{Gao201} defines discrete rewards for different user behaviors. Moreover, a one-step reward in \cite{Preda217} is distributed to the online recommendations during visitor interactions.

A creative work is reported in \cite{Chen12} which leverages GAN to model the dynamics of user behaviors and learn relevant reward functions, which depends not only on the user's historical behaviors but also on the selected item. The learned reward allows the agent to recommend items in a principled way, instead of relying on the manual design. Based on the simulation environment using the user behavior model, a cascading DQN is proposed to learn the combinatorial recommendation policy. Similarly, \cite{Chen187} introduces a generative IRL approach to avoid defining a reward function manually. The recommendation problem is regarded as automatic policy learning. Thus, this approach first generates a policy based on the user's preference. Later it uses a discriminative actor-critic network to evaluate the learned policy, based on the reward function defined by
\begin{equation}\label{equ:29}
  R(s, a) = log D(s, a) - log\big (max(\epsilon, 1-log D(s, a))\big )+r,
\end{equation}
where $ log D(s, a) $ is the negative reward generated by the discriminator, $r \in [0,1]$ represents the user’s feedback to prevent the agent from taking an extra step to update itself when the user clicks each recommendation, and $\epsilon$ is the maximum percentage of change that is updated at a time.


For commercial recommendation systems, online advertising is frequently inserted into personalized recommendation to maximize the profit. To this aim, a value-aware recommendation model \cite{Pei43} based on RL is designed to optimize the economic value of candidate items to make recommendations. In the value-aware model, measuring by the gross merchandise volume, the total reward is defined as the expected profit that is converted from all types of user actions (\eg click and pay). Moreover, \cite{Zhao99} presents an advertising strategy for DQN-based recommendation, where the advertising agent simultaneously maximizes the income of advertising and minimizes the negative influence of advertising on the user experience. Thus, the reward contains both the income of advertising and the influence of advertising on the user experience.

In practice, users have interests in exploring novel items. 
To this end, \cite{Zou44} proposes a fast Monte Carlo tree search method for diversifying recommendations, where the reward function is designed with the diversity and accuracy gain derived from recommending items at corresponding positions.

\subsection{Learning Bias}
RL algorithms can be classified into on-policy and off-policy methods \cite{Sutton72}. On-policy methods often sweep through all states to learn the policy and result in expensive costs, hence on-policy methods are not applicable to large-scale recommender systems. On the other hand, learning the recommendation policy from logged data (\eg logged user feedback) is a more practical solution, because it alleviates complex state space and high interaction cost with off-policy \cite{Precup120, Munos121}. Nevertheless, due to the difference between the target policy and the behavior policy, the off-policy methods usually result in data biases or policy biases.

\subsubsection{Data Biases}

There are lots of logged implicit feedback, such as user clicks and dwell time, available for learning users' preferences. However, the learning methods tend to easily suffer from biases caused by only observing partial feedback on previous recommendations. To deal with such data biases, \cite{Chen96} proposes a recommender system with the REINFORCE algorithm, where an off-policy correction approach is utilized to learn the recommendation policy from the logged implicit feedback, and incorporates the learned model of the behavior policy to adjust the data biases.

Differing from \cite{Chen96} that simply tackles data biases in the candidate generation module, the scalable recommender systems should contain not only the candidate generation stage but also a more powerful ranking stage. Toward this end, a two-stage off-policy method \cite{Ma97} is proposed to obtain data biases from logged user feedback and correct such biases by using inverse propensity scoring \cite{Horvitz122}. More precisely, the ranking module usually changes between the logged data (in behavior policy) and the candidate generation (in target policy). When there are evidently different preferences on the items to recommend, the two-stage off-policy policy gradient can be conducted to correct such biases. Moreover, the variance is reduced by introducing a hyper-parameter to down-weight the gradient of the sampled candidates.

\subsubsection{Policy Biases}
The direct offline learning methods, such as Monte Carlo and TD methods are subject to either huge computations or instability of convergence. To handle these problems, \cite{Zou90} proposes the PDQ framework to tackle the selection bias between the recommendation policy and logging policy. More specifically, the offline learning process is divided into two steps. In the first step, a user simulator is iteratively updated under the recommendation policy via de-biasing the selection bias. In the second step, the recommendation policy is improved with Q-learning, by both the logged data and user simulator. In addition, a regularizer for the reward function reduces both bias and variance in the learned policy.

Moreover, \cite{Bai23} proposes a model-based RL method to model the user-agent interactions for offline policy learning with adversarial training, where the agent interacts with the user behavior model to generate recommendations that approximate the true data distribution. To reduce the bias in the learned policy, the discriminator is employed to rescale the generated rewards for the policy learning, and de-bias the user behavior model by distinguishing simulated trajectories from the real interactions. As a result, the recommendation performance can be improved. Similarly, in the Adversarial future encoding (AFE) model \cite{Xie179}, a future-aware discriminator is taken as a recommendation module to identify user-item pairs, whereas a generator confuses the future-aware discriminator by generating items with only common features. The AFE model is optimized with the recommendation loss, which reduces the optimization bias in the pre-training task.

\subsection{Task Structuring}

The complexity of the RL task can generally be reduced by decomposing the RL task into several basic components or a sequence of subtasks \cite{Kober106}. For recommender systems, previous studies focus on Multi-Agent Reinforcement Learning (MARL), HRL, and Supervised Reinforcement Learning (SRL).

\subsubsection{Multi-agent Reinforcement Learning}
RL-based recommender systems suffer from inherent challenges (\eg the curse of dimensionality) when they employ a single agent to perform the task. An alternative solution is to leverage multiple agents with similar tasks to improve the learning efficiency with the help of parallel computation. Generally, MARL algorithms can be classified into four categories, \ie fully cooperative, fully competitive, both cooperative and competitive, and neither cooperative nor competitive tasks \cite{Busoniu134}.

Fully cooperative tasks in MARL-based recommender systems have the same goal (\ie maximizing the same cumulative return) achieved by all the agents\cite{Zhao223}. For instance, in the DEMER approach \cite{Shang16}, the policy agent cooperates with the environment agent to learn the policy of a hidden confounder. Moreover, to capture the general preference of users and their temporary interests, \cite{Du177} introduces a Temporary Interest Aware Recommendation (TIARec) model with MARL. Particularly, an auxiliary classifier agent can judge whether each interaction is atypical or not. The classifier agent and the recommender agent are jointly trained to maximize the cumulative return of the recommendations.

In contrast, fully competitive tasks typically adopt the mini-max principle: each agent maximizes its own benefit under the assumption that the opponents keep endeavoring to minimize it. In a dynamic collaboration recommendation method for recommending collaborators to scholars \cite{Zhang25}, the competition should be characterized as a latent factor, since scholars hope to compete for better candidates. To this end, the proposed method uses competitive MARL to model scholarly competition, \ie multiple agents (authors) compete with each other by learning the optimized recommendation trajectories. Besides that,
an improved market-based recommendation model \cite{Wei45} urges all agents to classify their recommendations into various internal quality levels, and employs Boltzmann exploration strategy to conduct these tasks by the recommender agents.

Both cooperative and competitive tasks also exist in MARL-based recommender systems\cite{Zhang211}. For example, \cite{Zhang186} aims to recommend public accessible charging stations intelligently. Each charging station is regarded as an individual agent. Subsequently, a centralized attentive critic module is developed to stimulate multiple agents to learn cooperative policies. While a delayed access strategy is proposed to exploit future charging competition information during model training.
Besides, to address the sub-optimal policy of ranking due to the competition between independent recommender modules, \cite{He22} promotes the cooperation of different modules by generating signals for these modules. Each agent acts on the basis of its signal without mutual communication. 
For the $i$-th agent with the signal vector $\phi^{i}$, given a Q-value function $ Q_{\theta}^{i}(S_{t},A_{t}) $ and a shared signal network $ \Phi^{i}(S_{t}) $, the objective function can be defined as follows.
\begin{align}\label{equ:30}
   J_{\phi}(\xi)= &\frac{1}{N}\sum_{i}\big[\mathbb{E}_{S_{t},A_{t}^{-i}\sim D,\phi^{i}\sim \Phi^{i}}[-Q_{\theta}^{i}(S_{t},A_{t}^{-i},\pi^{i}(S_{t},\phi_{t}^{i}))\nonumber\\
   &+\alpha log \Phi^{i}(\phi_{t}^{i}|S_{t})]\big],
\end{align}
where $ \xi $ and $ \theta $ are the network parameters, $ D $ denotes the distribution of samples, and $ N $ is the batch size. The objective function for each agent is optimized by the SAC algorithm.

\subsubsection{Hierarchical Reinforcement Learning}
In HRL, an RL problem can be decomposed into a hierarchy of subproblems or subtasks, which reduces the computational complexity. There exist some studies for HRL, which solve the recommendation problems well. For example, using Hierarchies of Abstract Machines (HAMs) \cite{Ronald103}, \cite{Zhang29} formalizes the overall task of profile reviser as an MDP, and decomposes the task MDP into two abstract subtasks. If the agent decides to revise the user profile (\ie a high-level task), it allows the high-level task to call a low-level task to remove noisy courses. 
To improve the recommendation adaptivity, a Dynamic Attention and hierarchical Reinforcement Learning (DARL) framework \cite{Lin133} is developed to automatically track the changes of the user's preferences in each interaction. These two methods adopt the REINFORCE algorithm to optimize both the high-level and low-level policy functions.

HRL with the MaxQ approach \cite{Thomas104} is a task-hierarchy that restricts subtasks to different subsets of states, actions, and policies of the task MDP without importing extra states. For recommender systems, a multi-goals abstraction based HRL \cite{Zhao30} is designed to learn the user's hierarchical interests. In addition, \cite{Xie98} proposes a novel HRL model for the integrated recommendation (\ie simultaneously recommending the heterogeneous items from different sources). In the proposed model, the task of integrated recommendation is divided into two subtasks (\ie sequentially recommending channels and items). The low-level agent works as a channel selector, which provides personalized channel lists in terms of user's preferences. On the other hand, the high-level agent is regarded as an item recommender, which recommends heterogeneous items based on the channel constraints.

Another prevailing approach is the options framework (\ie closed-loop policies for taking action over a period of time) \cite{Sutton105}, which generalizes the primitive actions to include a temporally extended navigation of actions. For example, \cite{Wang182} designs a Subgoal conditioned Hierarchical Imitation Learning (SHIL) framework for dynamic treatment recommendation. In the SHIL framework, the high-level policy sequentially selects a subgoal for each sub-task. Based on each subgoal, the low-level policy produces the low-level action (\ie effective medication) in the corresponding state for each sub-task.

\begin{table*}[t]
  \centering
  \caption{Overview of challenges in RL-based recommendation approaches.}
\begin{tabular}{llll}
\hline
\multicolumn{2}{l}{Issue}                                                        & Model                                                                                                                                                                                            & Evaluation Strategy        \\ \hline
\multirow{6}{*}{\begin{tabular}[c]{@{}l@{}}Environment \\ Construction\end{tabular}} & \multirow{2}{*}{State Representation} & DRR \cite{Liu02}, DEMER \cite{Sutton72}                                                                                                                                                           & Online \& Offline        \\
                                          &                                       & PyRecGym \cite{Bichen147}                   & Offline                  \\ \cline{2-4} 
                                          & KG Leverage              & \begin{tabular}[c]{@{}l@{}}KERL \cite{Wang10}, KGQR \cite{Zhou27}, PGPR \cite{Xian06}, ADAC \cite{Zhao57},\\ Attacks\&Detection \cite{Cao40}, KGPolicy \cite{Wang18}\end{tabular}                 & Offline                  \\ \cline{2-4} 
                                          & Negative Sampling                     & KGPolicy \cite{Wang18}, RNS \cite{Ding35}, SNQN \cite{Xin206}, DCFGAN \cite{Zhao200}                                         & Offline                  \\ \cline{2-4} 
                                          & \multirow{2}{*}{Social Relation}      & SADQN \cite{Lei48}                               & Online                   \\
                                          &                                       & SRRL \cite{Lu50}                                    & Offline                  \\ \hline
\multirow{1}{*}{Prior Knowledge}          & \multirow{1}{*}{}       & CBHR \cite{David55}, DR \cite{Liang52}, GAttN \cite{Gong28}                                                                                                                      & Offline           \\ \hline
\multirow{3}{*}{\begin{tabular}[c]{@{}l@{}}Reward Function \\ Definition\end{tabular}}        & \multirow{3}{*}{}                     & \begin{tabular}[c]{@{}l@{}}SQN and SAC \cite{Xin24}, VPQ \cite{Gao201}, GAN-CDQN \cite{Chen12} \end{tabular}                                                      & Offline                  \\
                                          &                                       & DEAR \cite{Zhao99}                                      & Online \& Offline        \\
                                          &                                       & PWR\cite{Preda217}, Value-based RL \cite{Pei43}, GIRL \cite{Chen187}                                 & Online                   \\ \hline
\multirow{3}{*}{Learning Bias}            & Data Biases                           & Off-policy Correction \cite{Chen96}, 2-IPS \cite{Ma97}                                                                                                                                            & Online \& Offline        \\ \cline{2-4} 
                                          & \multirow{2}{*}{Policy Biases}        & PDQ \cite{Zou90}                                 & Offline                  \\
                                          &                                       & IRecGAN \cite{Bai23}, AFE \cite{Xie179}                                   & Online \& Offline        \\ \hline
\multirow{6}{*}{Task Structuring}         & \multirow{2}{*}{Multi-agent RL}       & DEMER \cite{Shang16}, MASSA \cite{He22}, DeepChain \cite{Zhao223}, RLCharge \cite{Zhang211}                                                                                                                                                           & Online \& Offline        \\
                                          &                                       & Multi-with RL \cite{Zhang25}, INQ \cite{Wei45}, TIARec \cite{Du177}, MASTER \cite{Zhang186}                                                 & Offline                  \\ \cline{2-4} 
                                          & \multirow{2}{*}{Hierarchical RL}      & HRL+NAIS \cite{Zhang29}, DARL \cite{Lin133}, SHIL \cite{Wang182}                            & Offline                  \\
                                          &                                       & MaHRL \cite{Zhao30}, HRL-Rec \cite{Xie98}                                    & Online \& Offline        \\ \cline{2-4} 
                                          & \multirow{2}{*}{Supervised RL}        & \begin{tabular}[c]{@{}l@{}}SQN and SAC \cite{Xin24}, SRL-RNN \cite{Wang32}, SL+RL \cite{Huang58}, \\ PAR \cite{Georgios08}, Off-policy with guarantees \cite{Georgios62}\end{tabular} & Offline                  \\
                                          &                                       & SRR \cite{Liu49}, EDRR \cite{Liu146}                       & Online \& Offline        \\ \hline
\end{tabular}
\label{issues}
\end{table*}

\subsubsection{Supervised Reinforcement Learning}

In practice, the recommendation problems can be more easily solved by utilizing a combination of RL and supervised learning rather than only using RL, as supervised learning and RL can handle corresponding tasks according to their advantages respectively. For example, \cite{Xin24} proposes a self-supervised RL model to learn the recommendation policy from users' logged feedback in sequential recommender systems. The model has two output layers: One is the self-supervised layer trained with cross-entropy loss function to perform ranking.
The other is trained with RL based on a flexible reward function, which performs as a regularizer for the supervised layer. Studies in  \cite{Georgios08, Georgios62} focus on optimizing the user's life-time value in personalized ad recommender systems. To achieve this goal, they propose an RL-based recommendation model, where mapping from features to actions is learned by a random forest algorithm.

SRL is often applied to satisfy the adaptability of recommendation strategies. For instance, to address the top-aware drawback (\ie the performance on the top positions) that may reduce user experiences, a supervised DRL model \cite{Liu49} jointly utilizes the supervision signals and RL signal to learn the recommendation policy in a complementary way. Different from the top-aware recommender distillation framework \cite{Huang195} that utilizes DQN to reinforce the rank of recommendation lists, the supervised DRL model contains two styles of supervision signals. It adopts the cross-entropy loss for classification-based supervision signal, and employs pairwise ranking-based loss for the ranking-based supervision signal. The suitable supervision signals adaptively balance the long-term reward and immediate reward.

Moreover, \cite{Wang32} puts forward SRL with RNN for dynamic treatment recommender systems. By combining the supervised learning signal (\eg the indicator of doctor prescriptions) with the RL signal (\eg maximizing the cumulative reward from survival rates), this approach learns the prescription policy to refrain from unacceptable risks and provides the optimal treatment.
\cite{Huang58} also proposes a novel SRL with RNN for the long-term recommendation. More precisely, RNN is employed to adaptively evolve user states for simulating the sequential interactions between recommender system and users.

To tackle the training compatibility among the components of embedding, state representation, and policy in RL-based recommender systems,\cite{Liu146} proposes an End-to-end DRL-based Recommendation framework namely EDRR, in which a supervised learning signal is designed as a classifier of the user’s feedback on the recommendation results. Moreover, DQN and DDPG are employed to elaborate how embedding component works in the proposed framework.

\subsection{Summary}
An overview of challenges in RL-based recommendation approaches is presented in Table~\ref{issues}. Overall, a number of studies have attempted to address the challenges of applying RL in recommender systems. On the one hand, many studies focus on adapting RL algorithms to different recommendation scenarios. For instance, a promising reward function that is skillfully designed can meet specific requirements of the recommender system. On the other hand, some studies leverage related techniques to solve the issues of RL-based recommender systems. For example, the KG not only enriches user-item relations by the structural knowledge to describe the MDP environment but also contributes to explainable recommendations by the multi-hop paths. Nonetheless, there are some other possible limitations of RL-based recommender systems, as well as the challenges in complex application scenarios.
For these limitations and challenges, we provide detailed discussions in the following section.

\section{Discussion}
\label{Discussion}
In this section, we first discuss the central issues in applying RL algorithms for recommender systems. Afterwards, we analyze practical challenges and provide several potential insights into successful techniques for RL-based recommender systems.

\subsection{Open Issues}
In practice, RL algorithms are not applicable to real-time recommender systems, due to their trial-and-error mechanism. A key step in the application of RL algorithms is to improve the recommendation effectiveness and realize an intelligent design.
Hence, the following issues need to be solved for making much progress in this field.

\subsubsection{Sampling Efficiency}
Sampling plays a substantial role in RL, especially in DRL-based recommender systems. Importance sampling has empirically demonstrated its application feasibility in both on-policy and off-policy. Some works \cite{Wang18, Ding35} have highlighted the superiority of negative sampling in recommender systems. Nevertheless, it is necessary to focus on the improvement in sampling efficiency, since the user feedback available to train recommendation models is scarce. We can leverage auxiliary tasks to improve sampling efficiency. For example, \cite{Minmin157} develops a user response model to predict the user's positive or negative responses towards the recommendation results. Thus, the state and action representations can be enhanced with these responses. Moreover, transfer learning may be qualified for the sampling task. For example, we can use the transfer of knowledge between temporal tasks \cite{Xu139} to perform RL on an extended state space, and concretize similarity between the source task and the target task by logical transferability. In addition, the transfer of experience samples \cite{Andrea138} can estimate the relevance of source samples. Both transfer learning methods may be able to improve the sampling efficiency in RL-based recommender systems.

\subsubsection{Reproducibility}
In many application scenarios, including recommendation, it is often difficult to reproduce the results of RL algorithms, because of various factors such as instability of RL algorithms, lacking open source codes \& account of hyper-parameters, and different simulation environments (\eg experimental setup and implementation details), although most existing studies have performed empirical experiments on public datasets, as shown in Table~\ref{TRS}. Especially for DRL methods, non-determinism in policy network and intrinsic variance of these methods leads to reproducibility crisis.
Besides policy evaluation methods, there are several future directions for the reproducibility investigation with statistical analysis. 
We can use significance testing according to related metrics or hyper-parameters such as batch size and reward scale \cite{Peter165}. For example, the reward scale needs to check its rationality and robustness in specific recommendation scenarios. The average returns should be evaluated to verify whether it is proportional to the relevant performance. Moreover, due to the sensitivity of RL algorithms changed in their environments, random seeds, and definition of the reward function, we should ensure reproducible results with fair comparisons. To achieve this objective,  we need to run the same experiment trials for baseline algorithms, and take each evaluation with the same preset epoch. Moreover, all experiments should adopt the same random seeds \cite{Peter165}.

\subsubsection{MDP Formulation}
In principle, MDP formulation is essential to guarantee the performance of RL algorithms. However, the state and action spaces in recommender systems often suffer from the curse of dimensionality. Besides, the reward definition is sensitive to the external environment since users' demands may change constantly, while interactions between users and items are random or uncertain. To address these issues, we can employ task-specific inductive biases to learn the representations of selection-specific action. As a result, the agent relies on better action structures to learn RL policies when the recommendation problem is formulated \cite{Welborn174}. Another direction is to use causal graph and probabilistic graph models \cite{Xu175} to enable the MDP formulation, where the causal graph describes the causal-effect relations among user-item interactions, and the probabilistic graph reasons the recommendation paths. This joint process can improve the efficiency of agent search and tracks users' interests over time. However, how to design MDPs for recommendation problems in a verifiable way, remains an open issue.

\subsubsection{Generalization}
Model generalization ability is almost the pursuit of all applications. Limited by the shortcomings of different RL algorithms, it is difficult to develop a general framework to meet various specific requirements of recommendation. On the one hand, model-based algorithms are only applicable to the specified recommendation problem, and fail to solve the problem that cannot be modeled. On the other hand, model-free algorithms are insufficient in dealing with different tasks in complex environments.
Fortunately, meta-RL, such as Meta-Strategy \cite{Powers135} and Meta-Q-Learning \cite{Fakoor140}, first learns a large number of RL tasks to obtain enough prior knowledge, and later can be quickly adapted to the new environment in face of new tasks. In this case, it may enable the generalization of RL-based recommender systems. Besides transfer learning based on meta-RL, we can use multi-task learning\cite{Zhang215} to handle related tasks (\eg construction of user profiles, recommendation, and causal reasoning) in parallel. These tasks complement each other to improve the generalization performance via shared representation of domain information, such as parameter-based sharing, and joint feature-based sharing.

\subsubsection{Autonomy}
It seems easy to utilize RL for autonomous control \cite{Kiumarsi196}. However, in practice, it is difficult to achieve this objective in online recommendation scenarios, since there are complex interactions between users and items, while it is not feasible to capture users' dynamic intentions by existing RL algorithms. To this end, we can achieve the feedback control of recommender systems by combining RL with LSTM or GRU, which enables the RL agent to be possessed of powerful memory that preserves the sequences of state and action, as well as different kinds of reward. Later, the historical information is encoded and transmitted to the policy network, thereby improving the autonomous navigation ability of the RL agent. Another interesting research direction is to improve the autonomous learning ability of RL from the fields of biology, neuroscience, and cybernetics. Thus, it is straightforward to use the learned knowledge to make better recommendations.

\subsection{Practical Challenges}

The existing recommendation models based on RL have demonstrated superior recommendation performance. Nevertheless, there are many challenges and opportunities in this field. We summarize five potential directions that deserve more research efforts from the practical aspect.

\subsubsection{Computational Complexity}
RL-Based recommender systems often suffer from huge computations due to the exploration-exploitation tradeoff and the curse of dimensionality \cite{Xie188}. Apart from task structuring techniques, the IRL algorithm is an efficient solution to reduce the computation cost. Initializing RL with demonstrations can supervise the agent taking the action correctly, particularly for some specific recommendation tasks. For example, apprenticeship learning via IRL algorithm \cite{Abbeel109} highlights the need for learning from an expert, which maximizes a reward based on a linear combination of known features. 
The hierarchical DQNs \cite{Fu191} alleviate the curse of dimensionality in large-scale recommender systems. Besides, a promising method of driving route recommendation based on RL is to employ behavior cloning \cite{Torabi108}, which makes expert trajectories available and quickens the learning speed. Another feasible scheme is to improve the efficiency of agent exploration. For example, we can adopt NoisyNet \cite{Meire166} that adds parametric noise to its network weights, which aids efficient exploration according to the stochasticity of the policy for the recommender agent. 

\subsubsection{Evaluation}
Most existing RL-based recommender systems focus on the single goal of recommendation accuracy, without considering recommendation novelty and diversity \cite{Kunaver116} based on the user experience. Beyond the need for new evaluation measures for multi-objective goals of recommendation (\eg popularity rate \cite{Ge202}), we should design standard metrics for other non-standardized evaluation measures (\eg diversity, novelty, explainability, and safety). In general, these kinds of evaluation measures can be considered to be combinatorial optimization problems, naturally, maybe well achieved by multi-objective evolutionary algorithms \cite{Christian142}. Recently, some works have developed Pareto efficiency models for multi-objective recommendation \cite{Chen169, Dusan207}. For example, in the personalized approximate Pareto-efficient recommender system \cite{Xie170}, a Pareto-oriented RL module learns personalized objective weights on multiple objectives for the target user. Nevertheless, it remains a challenge to reconcile different evaluation measures for the recommendation, since these evaluation measures are usually relevant and even conflict.

\subsubsection{Biases}
In recommender systems, item popularity often changes over time due to the user engagement and recommendation strategy \cite{Ge156}, thus long-tailed items are rarely recommended to users. The selection bias may hurt user satisfaction\cite{Li214}. Therefore, it is necessary to concentrate on the fairness work. Towards this end, we can combine anthropology to analyze the differences in human behavior and cultural characteristics, or explore the user's intentions in the user-recommender interactions. Besides, heterogeneous data of the user behavior often exists in online recommendation platforms, whereas most recommendation approaches are trained with a single type of data. Due to the information asymmetry between the user behavior and training model, the recommender system suffers from learning bias. Undoubtedly, the previous feature extraction and representation learning are crucial to deal with such bias. In addition, MARL can be used to encourage different agents to execute multi-dimensional data simultaneously, and share parameters for a unified recommendation goal.

\subsubsection{Interpretability}
Due to the complexity of RL, the post-hoc explanation may be easier to achieve than intrinsic interpretability \cite{Puiutta136}. Actually, the same is true in RL-based recommendation systems. How to explore intrinsic interpretable methods for RL to provide more convincing recommendations is a promising line. Another research direction towards explainable recommendation is to provide formal guidance of recommendation reasoning process, rather than being concerned with a multi-hop reasoning process\cite{Wu209}. \cite{Tai168} proposes a user-centric path reasoning framework that adopts a multi-view structure to combine sequence reasoning information with subsequent user decision-making. However, they only focus on the user's demand and do not provide theoretical proof of the rationality of reasoning. We should avoid plausible explanations in the reasoning process. To this end, we can employ multi-task learning to perform the recommendation reasoning process among multiple related tasks, \eg representation of interaction, recommended path generation, and Bayesian inference for policy network. These related tasks jointly provide credible recommendation explanations by sharing the representation of internal correlation and causality.

\subsubsection{Safety and Privacy}
System security and user privacy are important issues, which are ignored by most existing studies. For example, personal privacy is easy to be leaked when using RL and KG to perform the explainable recommendation, because the relationships among users and items are exposed. Differential privacy is widely applied to protect user privacy, and DRL can be employed to choose the privacy budget against inference attacks \cite{Xiao155}. Besides, recent studies have found that Deep Neural Networks (DNNs) are vulnerable to attacks, such as adversarial attacks and data poisoning. For example, in DNNs-based recommender systems, users with fake profiles may be generated to promote selected items. To address this issue, a novel black-box attacking framework \cite{Fan171} adopts policy gradient networks to refine real user profiles from a source domain and copy them into the target domain. 
Notwithstanding, we need to make more efforts on the research topic. For example, the safe RL \cite{Javier137} can be utilized to guarantee the security of recommender systems (\eg detecting error states, and preventing abnormal agent actions). We may also leverage federated learning \cite{Li143, Huang192} to achieve privacy-preserving data analysis for MARL-based recommender systems, where each agent is localized on distributed device and updates a local model based on the user data stored in the corresponding user device.

\section{Conclusion}
\label{Conclusion}
Recommender systems serve as a powerful technique to address the information overload issue on the Internet. There has been increasing interest in extending RL approaches for recommendations in recent years. RL-based recommendation methods autonomously learn the optimal recommendation policies from user-item interactions, and thus they recommend better items to users, compared with other recommendation methods. In this survey, we firstly conduct a comprehensive review on RL-based recommender systems, using three major categories of RL (\ie value-function, policy search, and Actor-Critic) to cover four typical recommendation scenarios. We also restructure the general frameworks for some specific scenarios, such as interactive recommendation, conversational recommendation, and the explainable recommendation based on KG. Furthermore, the challenges of applying RL in recommender systems are systematically analyzed, including environment construction, prior knowledge, the definition of the reward function, learning bias, and task structuring.
To facilitate future progress in this field, we discuss theoretical issues of RL and analyze the limitations of existing approaches, and finally put forward some possible future directions.






\vspace{11pt}


\bibliographystyle{IEEEtran}
\bibliography{IEEEabrv}

\vfill

\end{document}